\documentclass{emulateapj}
\usepackage{natbib}
\def\msun{M$_{\odot}$}
\def\s2n{S^{\prime}/N}

\graphicspath{{figures/}}

\shorttitle{}
\shortauthors{Padoan, Haugb{\o}lle, Nordlund}

\begin{document}
\title{Infall-Driven Protostellar Accretion and the Solution to the Luminosity Problem}

\author{Paolo Padoan}
\affil{ICREA \& Institut de Ci\`{e}ncies del Cosmos, Universitat de Barcelona, IEEC-UB, Mart\'{i} Franqu\`{e}s 1, E08028 Barcelona, Spain; ppadoan@icc.ub.edu}
\author{Troels Haugb{\o}lle}
\affil{Center for Star and Planet Formation, Natural History Museum of Denmark, University of Copenhagen, {\O}ster Voldgade 5-7, DK-1350 Copenhagen K, Denmark}
\affil{Center for star and Planet Formation, Niels Bohr Institute, University of Copenhagen, Juliane Maries Vej 30, DK-2100 Copenhagen, Denmark}
\author{{\AA}ke Nordlund}
\affil{Center for star and Planet Formation, Niels Bohr Institute, University of Copenhagen, Juliane Maries Vej 30, DK-2100 Copenhagen, Denmark}
\affil{Center for Star and Planet Formation, Natural History Museum of Denmark, University of Copenhagen, {\O}ster Voldgade 5-7, DK-1350 Copenhagen K, Denmark}

\begin{abstract}

We investigate the role of mass infall in the formation and evolution of protostars. 
To avoid ad hoc initial and boundary conditions, 
we consider the infall resulting self-consistently from modeling the formation of stellar clusters in turbulent molecular clouds. 
We show that infall rates in turbulent clouds are comparable to accretion rates inferred from protostellar luminosities or measured in 
pre-main-sequence stars. They should not be neglected in modeling the luminosity of protostars and the evolution of disks, even after the embedded protostellar phase.
We find large variations of infall rates from protostar to protostar, and large fluctuations during the evolution of individuals protostars.
In most cases, the infall rate is initially of order 10$^{-5}$~\msun\ yr$^{-1}$, and may either decay rapidly in the formation of low-mass
stars, or remain relatively large when more massive stars are formed. The simulation reproduces well the observed characteristic values and scatter 
of protostellar luminosities and matches the observed protostellar luminosity function. The luminosity problem is therefore solved once 
realistic protostellar infall histories are accounted for, with no need for extreme accretion episodes. These results are based on a
simulation of randomly-driven magneto-hydrodynamic turbulence on a scale of 4~pc, including self-gravity, adaptive-mesh refinement
to a resolution of 50~AU, and accreting sink particles. The simulation yields a low star formation rate, consistent with the observations, and a 
mass distribution of sink particles consistent with the observed stellar initial mass function during the whole duration of the simulation, forming
nearly 1,300 sink particles over 3.2 Myr.

\end{abstract}


\section{Introduction}

The study of the initial growth of protostars has important implications for our theoretical understanding of star and 
planet formation and for a correct interpretation of observations and of isotopic abundances in meteorites. 
Idealized models for the formation of individual stars, where the initial condition is an isolated core, 
yield specific predictions for the time evolution of the infall rate and accretion luminosity. In the case of 
embedded protostars, where the accretion energy is the main luminosity source, infall models yield much too large
luminosities compared to the observations, and cannot explain the observed luminosity scatter covering at least two orders 
of magnitude \citep[e.g.][]{Evans+09}. This is known as the luminosity problem, first noticed by \citet{Kenyon+90}.
The infall rate, through its effect on disk instabilities and chemical evolution, must also be crucial in modeling planet formation. 
Furthermore, the isotopic composition of various components of chondritic meteorites, particularly calcium-aluminum inclusions 
and chondrules, provides important clues about the early evolution of the solar nebula that can only be interpreted in the context 
of a holistic disk model where the infall rate may play an important role \citep[e.g.][]{Connelly+2012}.

The process of star formation is conventionally portrayed as composed of two main stages: i) The gravitational collapse of the 
protostellar core, when the central object acquires most of its final mass and is referred to as a protostar; ii) the pre-main-sequence 
(PMS) contraction, when infall from the collapsing core has subsided and the central object is referred to as a PMS star. Using the 
infrared spectral index classification of \citet{Lada+Wilking84}, the observational counterparts of protostars are Class 0 and I sources, 
while Class II and III are PMS stars. It is generally assumed that both protostars and PMS stars increase their mass through disk accretion. 
However, accretion rates from the disk to the stellar surface can be measured only in PMS stars \cite[e.g.][]{Manara+2012,Manara+2013a,Alcala+2014},
where they are used to constrain models of disk evolution \citep[e.g.][]{Sicilia-Aguilar+2010,Bae+2013,DaRio+2014,Ercolano+2014}. 
In embedded protostars, neither the infall rate from the collapsing core, nor the accretion rate from the disk to the star can be easily measured,
and the accretion rate can only be estimated from the total protostellar luminosity and a fair amount of modeling and assumptions. 

Models of the evolution of the protostellar luminosity tend to focus on the role of disk accretion, particularly in the case of episodic accretion, 
one of the most studied solution to the luminosity problem \citep[e.g.][]{Zhu+2010,Zhu+2010b,Baraffe+2012,Dunham+Vorobyov2012,Audard+2014},
while the infall is simply assumed to be that of an isolated collapsing core with very specific and idealized initial conditions. In the case of PMS
stars, infall is usually completely ignored, and the disks are assumed to evolve in isolation. Although the disk is certainly a necessary channel
for mass accretion, we contend that focusing on disk physics, while glossing over important aspects of larger-scale mass infall, may not be the 
best way to pursue a quantitative description of the accretion rate. We also argue that the infall should not be neglected in the study of PMS stars. 

We propose a new paradigm where the accretion rate is primarily controlled by the mass infall from larger scales (not just from an isolated protostellar core), 
for both protostars and young
PMS stars (Class II and III), while disk physics modulates, but does not control, the accretion. Although the infall rate onto the disk
has been accounted for in studies of protostellar luminosity cited above, it has been modeled as the result of the gravitational collapse of a highly idealized, isolated core,
with ad hoc initial conditions. The resulting infall rate is completely dependent on the adopted initial conditions. The process of core formation
has been neglected, including the role of converging flows feeding the core from larger scales. In our approach, we avoid 
using ad hoc initial and boundary conditions, and instead pursue a very realistic description of the infall rates, consistent with the large-scale 
dynamics and capable of reproducing the correct stellar initial mass function (IMF), as well as a realistic star formation rate (SFR). In the context of modeling
the protostellar luminosity, it is thus the 
first time that the role of the infall rate is accounted for in a self-consistent way, as well as the first time that it is quantified past the embedded phase
(following the idea suggested in \citet{Padoan+2005_accr}). We achieve this by modeling ab initio the birth and evolution of over one thousand
protostars, running a simulation of a relatively large turbulent region, of approximately 4~pc, with average properties typical of observed 
molecular clouds (MCs), for 3.2 Myr.

In the scenario of turbulent fragmentation \citep[e.g.][]{Larson81,Elmegreen93,Padoan95,Klessen+2000,Padoan+2001cores,Heitsch2001,Klessen2001,
Padoan+Nordlund02imf,Tilley+Pudritz2004,Clark+Bonnell2005,Klessen+2005,Padoan+2013ppvi}, 
protostellar cores are the natural outcome of converging flows in turbulent clouds \citep{Elmegreen93,Padoan+2001cores}. 
Due to the stochastic nature of turbulent flows, infall rates feeding the core from relatively large scales can be highly variable in time and space. 
Once a core reaches a critical mass for gravitational instability it collapses into a protostar. However, the core mass at that stage is not a tight constraint 
on the final stellar mass, because the infall rate is controlled by converging motions in the turbulent flow that can have 
a significantly longer timescale than the initial free-fall time of the core \citep{Padoan+Nordlund2011imf}. Generally speaking, infall rates of longer duration 
and/or higher values are required to form more massive stars. 

Disks are the necessary pathway for gas accretion onto the 
protostellar surface, but, as clearly suggested by their low mass, they cannot serve as the main mass reservoirs feeding the growth of protostars.
The disk-to-star mass ratio is typically in the range 0.2--0.6\% in Class II sources \citep[][]{Andrews+2013}, and $<0.1$ in Class I and Class 0 protostars 
\citep{Jorgensen+2009,Choi+2010,Chiang+2012,Tobin+2013,Murillo+2013,Harsono+2014,Lindberg+2014,Miotello+2014}, though possibly $>0.1$ in some Class 0 protostars 
\citep{Harsono+2014}. The main mass reservoir and driver of protostellar growth must be the infall of gas from the initial protostellar core collapse and from the same 
converging flows that formed that protostellar core. Even once a protostar has left its original birth site, a converging region in the turbulent flow, and has acquired most 
of its final mass, turning into a PMS star, Bondi-Hoyle accretion \citep[e.g.][]{Edgar2004,Ruffert97} can still be comparable to the observed accretion rates 
\citep{Padoan+2005_accr,Throop+Bally2008,Scicluna+2014}.

Our idea that disks are not the main (isolated) mass reservoir even for young PMS stars, is also
supported by the observational evidence of grain growth. The dust opacity coefficient of protostellar disks is found to be on average $\beta\approx 0.5$, 
much lower than the typical ISM value, which is interpreted as evidence for rapid grain growth up to mm size in disks \citep{Testi+2014ppvi}. 
If disks evolved in relative isolation, their dust content would also continuously evolve, with the largest grains being gradually lost by radial drift. 
Nevertheless, the opacity coefficient shows no time evolution \citep{Ricci+2010oph,Ricci+2010ta}, which we regard as suggestive of a 
continuous replenishment of the disk dust and gas through infall of fresh material, approximately balancing the accretion rate. The same conclusion
may be reached from the very similar distributions of silicate feature characteristics in Spitzer disk sources from regions with different median ages
\citep{Oliveira+2010}.

In this work, we address the problem of the formation and growth of protostars by studying the infall rate over a period of time continuing well beyond the 
embedded phase. We do not model the internal disk processes that make the accretion possible, but assume that the infalling mass finds
its way to the protostar or the PMS star, irrespective of the specific processes allowing this to occur. This assumption is supported by our finding 
that the infall rates we predict are consistent with the inferred accretion rates of protostars and the observed accretion rates of PMS stars.

\begin{figure*}[t]
\includegraphics[width=18cm]{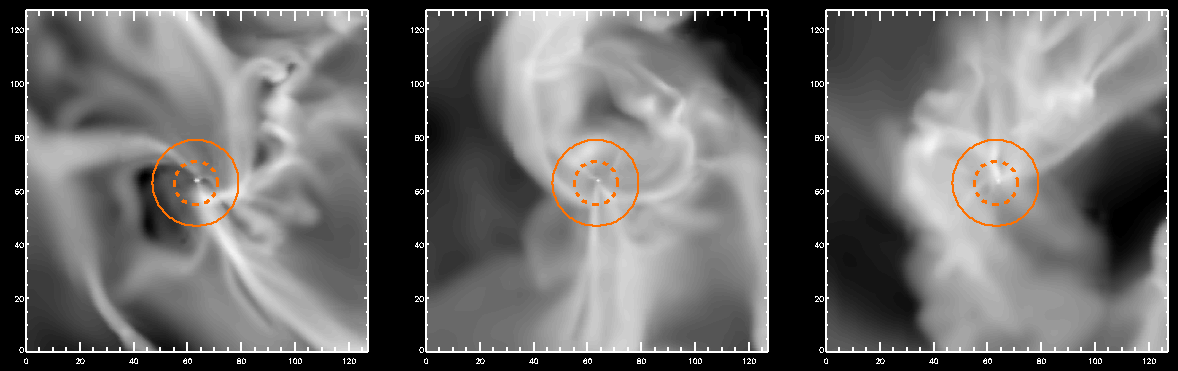}
\includegraphics[width=18cm]{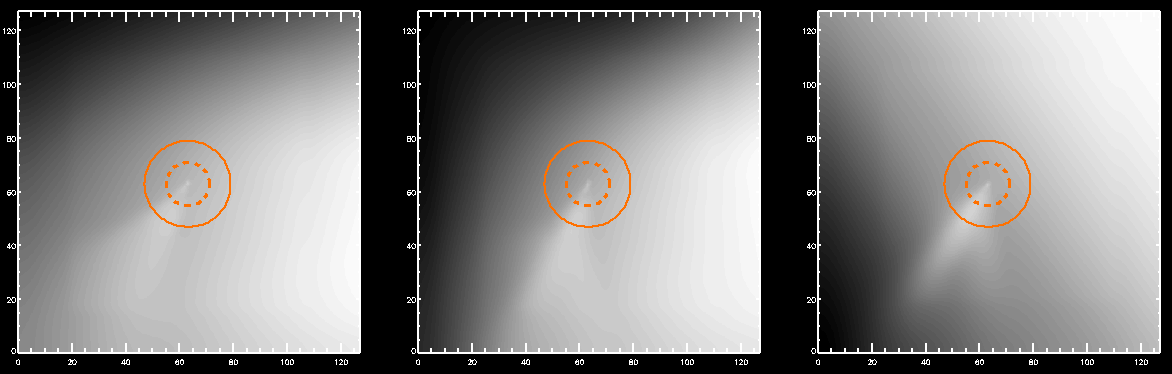}
\caption{Density projection of two regions of 6400 AU size centered around two sink particles. The top row of panels show the three orthogonal 
direction of projections for a relatively young sink, embedded in dense gas and  undergoing a high accretion rate. The bottom panels show the same three
projections for an older sink, moving through a relatively low density region and experiencing a very low accretion rate. The two orange circles show the size
of the accretion region (dashed line) and of the exclusion region (solid line), with radii of  $r_{\rm accr}=8\, \Delta x=400$~AU, and $r_{\rm excl}=16\,\Delta x=800$~AU.}
\label{sinks_maps}
\end{figure*}

\section{The Simulation}

In this section we present the simulation, giving only a brief description of technical aspects of the code and of the sink particle
implementation. A more complete discussion of our sink particle implementation is given in Haugb{\o}lle et al. (2014), 
where we present the most extensive numerical study to date of the stellar initial mass function (IMF).     

The simulation was carried out using the public adaptive-mesh-refinement (AMR) code Ramses \citep{Teyssier02}, modified to 
include random turbulence driving, a novel algorithm for sink particles, and an improved HLLD solver to allow numerical stability 
in the high-Mach number regime. It required approximately one million CPU hours on the NASA/Ames Pleiades supercomputer. 
As in \citet{Padoan+Nordlund02imf,Padoan+Nordlund04bd,Padoan+Nordlund11sfr} and in \citet{Padoan+2012sfr}, we adopted periodic boundary 
conditions, an isothermal equation of state, and solenoidal random forcing in Fourier space at wavenumbers $1\le k\le2$ ($k=1$ 
corresponding to the computational box size). We chose a solenoidal force to guarantee that collapsing regions are naturally 
generated in the turbulent flow, rather than directly imposed by the external force. This driving force keeps the three-dimensional 
rms sonic Mach number, ${\cal M}_{\rm s}\equiv \sigma_{\rm v,3D}/c_{\rm s}$ ($\sigma_{\rm v,3D}$ is the three-dimensional rms 
velocity, and $c_{\rm s}$ is the isothermal speed of sound), at the approximate value of 10, characteristic of MCs on the scale of few pc. 

We solve the compressible MHD equations, without explicit viscosity or resistivity, starting from uniform gas density and magnetic field,
and zero velocity. Gravity is not included during the first 20 dynamical times, $t_{\rm dyn }$, where $t_{\rm dyn}\equiv L_{\rm box}/ (2 \sigma_{\rm v,3D})$,
so the turbulent flow can reach a statistical steady state, and the magnetic energy can be amplified to its saturation level. The initial
value of the uniform magnetic field was such that the rms Alfv\'{e}nic Mach number defined with respect to the (conserved) mean magnetic field is
${\cal M}_{\rm a}\approx 5$, where ${\cal M}_{\rm a}\equiv \sigma_{\rm v,3D}/v_{\rm a}$, and $v_{\rm a}$ is the Alfv\'{e}n velocity corresponding
to the mean magnetic field strength, $B_0$, and the mean density, $\rho_0$, $v_{\rm a}=B_0/\sqrt{4\,\pi \rho_0}$. 
The initial magnetic energy is readily amplified by stretching and compression events in the turbulent flow, so it is important to 
run the simulation for several dynamical times until a saturation level has been reached \citep{Federrath+2011}.

\begin{table*}\footnotesize
\caption{Non-dimensional and main physical parameters of the simulation.}
\centering
\begin{tabular}{cccccccccccc}
\hline\hline \\[-2.ex]
${\cal M}_{\rm s}$ & ${\cal M}_{\rm a}$ & $t_{\rm ff}/t_{\rm dyn}$ & $\alpha_{\rm vir}$  & $L_{\rm J}/\Delta x$  &  $T$  [K] & $L_{\rm box}$  [pc]   & $\Delta x$ [AU]  & $M_{\rm box}$  [\msun]  & $B_0$  [$\mu$G]  & $t_{\rm ff}$  [Myr]  & $t_{\rm dyn}$  [Myr] \\ [0.8ex]
\hline \\ [-2.ex]
10 &  5    &  1.13  &   0.83  &   14.4  &  10.0  &  4.0 & 50.0 &  2998 &  7.2  &  1.22  & 1.08        \\[0.8ex]
\hline \\ 
\end{tabular}
\label{t1}
\end{table*}

After gravity is included, the simulation is continued for three more dynamical times, at which point 1,288 sink particles have been created, with a total
mass of 16\% of the total initial gas mass, meaning that the final star formation efficiency (defined as the total mass in sink particles divided by the initial gas mass) 
is SFE=0.16.  The assumed strength of gravity is such that
the virial parameter, using a practical definition of $\alpha_{\rm vir} \equiv (5/6) \sigma_{\rm v,3D}^2 L_{\rm box} / (G M_{\rm box})$ \citep{Bertoldi+McKee92}, is 
$\alpha_{\rm vir} = 0.83$. This parameter expresses the ratio between thermal plus turbulent kinetic energy and gravitational energy, in the case of a uniform 
isothermal sphere. Its application as an approximate estimate of such energy ratio in simulations in non-trivial, both because
of the shape and boundary conditions of the numerical box, and because of the strong fragmentation in the turbulent gas \citep{Federrath+Klessen2012}.
For a more straightforward non-dimensional ratio, equivalent to $\alpha_{\rm vir}$, we also refer to the ratio of 
the free-fall time, $t_{\rm ff}\equiv \sqrt{3\pi/ (32 G \rho_0)}$, and the dynamical time, $t_{\rm dyn}\equiv L_{\rm box}/ (2 \sigma_{\rm v,3D})$ \citep{Padoan+2012sfr}, 
which in our simulation is $t_{\rm ff}/t_{\rm dyn} \approx 1.13$.

To scale the simulation to physical units, we adopt a temperature of $T=10$~K and a size of $L_{\rm box}=4$~pc, 
yielding $\sigma_{\rm v,3D} \approx 1.8$~km s$^{-1}$ (consistent with observed line width-size relations),
$M_{\rm box}\approx 2,998$~\msun\,, a mean number density of $n_0\approx 795$~cm$^{-3}$ (assuming a mean molecular weight of 2.4), a mean
magnetic field of $B_0=7.2$~$\mu$G, a dynamical time of $t_{\rm dyn}\approx 1.08$~Myr, and a free-fall time of $t_{\rm ff} \approx 1.22$ Myr. 
Thus, in physical units, the simulation is run with self-gravity for a period
of 3.2~Myr, comparable to the estimated age of many nearby young star-forming regions. As shown below, this is also a long enough time to allow for the 
formation of stars of a few solar masses and thus to accurately sample the Salpeter range of the stellar IMF. 
The fundamental non-dimensional parameters of the simulation, and the assumed values of the physical parameters are summarized in Table~1.

The root grid of this AMR simulation contains 256$^3$ computational cells, thus the minimum spatial resolution (in the lowest density regions) is 
$\Delta x_{\rm root}=4$~pc$/256=0.0156$~pc. We use 6 AMR levels, each increasing the spatial resolution by a factor of 2, thus our maximum spatial resolution
(in dense regions) is $\Delta x=50$~AU. The refinement criterion is based only on density: wherever the density on the root grid is larger than 10 times the mean
density, we add one refinement level (increase the resolution by a factor of two), and further AMR levels are added for each increase in density by a factor of 
4, in order to keep the shortest Jeans length equally refined at all levels. For the physical parameters given above, the Jeans length is always very well resolved,
$L_{\rm J}\ge 14.36\,\Delta x$ at every AMR level, except at the highest resolution, where we allow the density to grow by an extra factor of 4 before creating a sink particle,
in order to let the collapse evolve as long as possible. The gas number density at sink particle creation is thus $n_{\rm max}=3.3\times10^7$~cm$^{-3}$,
where the Jeans length is still relatively well resolved with $7.2\,\Delta x$ 
 
Details of the sink particle creation will be discussed in a separate paper (Haugb{\o}lle et al. 2014). Here, we just mention that a
sink particle is created in a cell of the highest resolution, where the gas density is $n \ge n_{\rm max}$, corresponding to $L_{\rm J}\le 7.2\,\Delta x$, as mentioned above. 
The creation of a sink particle also requires that the gravitational potential has a local minimum in the cell, and that the velocity field is converging in the cell, 
$\nabla \cdot {\bf u} < 0$. Furthermore, no other previously created sink particle can be present within an exclusion radius, $r_{\rm excl}=16\,\Delta x$, of the cell where 
the new particle is created. These conditions for sink particle creation are similar to those implemented in previous works 
\citep{Bate+95sink,Krumholz+04sink,Federrath+10sinks,Gong+Ostriker2013}.

\begin{figure*}[t]
\includegraphics[width=9cm]{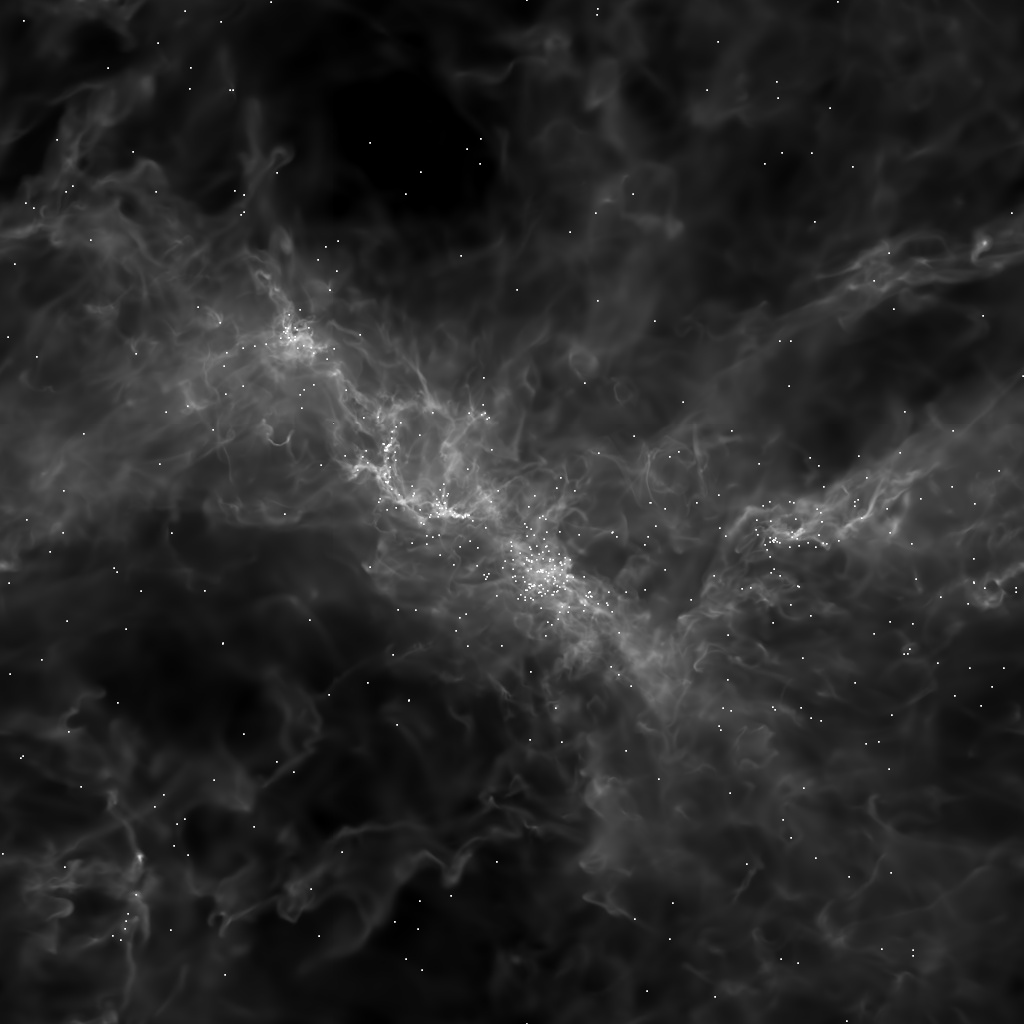}
\includegraphics[width=9cm]{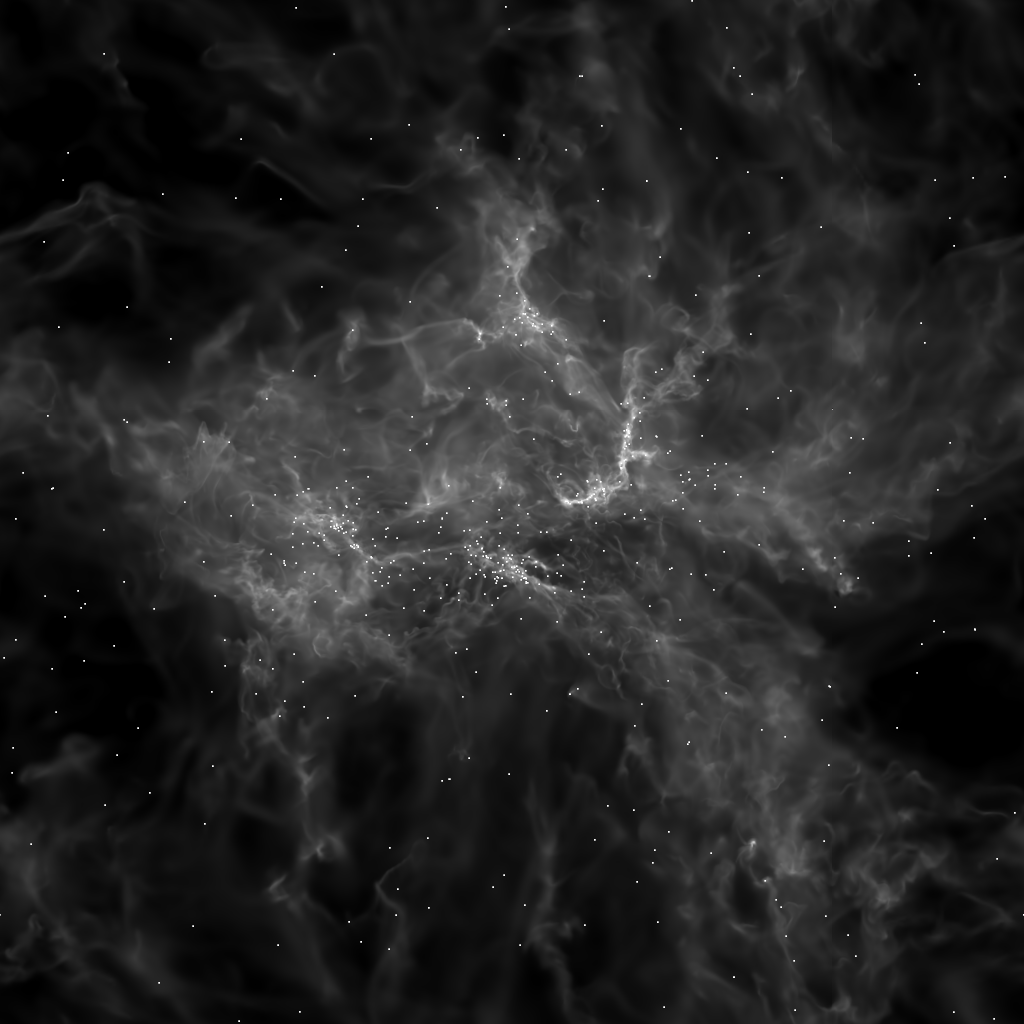}
\caption{Gas density of the simulation projected along the x axes (left panel) and the y axes (right panel), and position of all sink particles, at $t=2.6$~Myr 
after the formation of the first sink particle. The images cover the whole simulation, so they correspond to an area of $4\times 4$ pc$^2$. Because of the 
periodic boundary conditions, they have been shifted to center them around the densest cloud regions.}
\label{maps}
\end{figure*}

A sink particle is first created without any mass, but is immediately allowed to accrete. A sink particle in this simulation typically starts 
with a mass of $4\times 10^{-5}$~\msun . It accretes from cells that are closer than an accretion radius of $r_{\rm accr}=8\,\Delta x=400$~AU, as long as the 
gas in such cells is gravitationally bound to the sink and has a density larger than $n_{\rm accr}=10^{-3}n_{\rm max}$. Within the accretion radius,
the rate of accretion per unit mass varies smoothly, from zero at the edge, to  $\sim 0.1$ per orbital time near the sink particle.  Deeper zoom-in simulations,
with cell sizes down to a fraction of an AU \citep{Nordlund+2014}, have shown this to be appropriate, and, when applied at the current scales, it 
gives a good compromise between creating either artificial voids (too large accretion rate), or artificial mass accumulation (too low accretion rate) near the sinks.
Only a fraction $\epsilon_{\rm sink}=0.5$ of this accreting gas is given to the sink particle, to mimic the mass loss due to winds and jets. The other 
half of the accreting gas mass is simply removed from the simulation, without any feedback. 

Figure~\ref{sinks_maps} shows the projections of small volumes of $128^3$ cells of the highest resolution ($\Delta x=50$~AU), extracted around a very young sink 
(upper panels), and an older one (lower panel), along the x, y and z axis (left to right panels). The physical size of the images is thus $6400\times 6400$~AU. The 
accretion and exclusion regions are also shown. One can see that our choice of accretion parameters is such that the gas dynamics within the accretion radius is not 
unphysically perturbed (filaments cross the accretion radius without exhibiting numerical artifacts), thanks to our choice of accretion rate. The images also illustrate 
that young sink particles are fed by filamentary infall from larger scales, while older ones are fed by Bondi-Hoyle accretion, with characteristic hollow shock cones
downstream of the sink particles \citep[e.g.][]{Ruffert97,Ruffert99}.

With such parameters, we can in principle detect infall rates as low as $\sim10^{-16}$~\msun\ yr$^{-1}$, but infall rates are not smoothly resolved below  
$\sim 10^{-11}$~\msun\ yr$^{-1}$. In order to smoothly resolve infall rates of $\sim 10^{-14}$~\msun\ yr$^{-1}$ we have rerun a few stretches of the simulation 
with a 1,000 times lower accretion density limit, $n_{\rm accr}=10^{-6}n_{\rm max}=33$~cm$^{-3}$. All our results (except for plots showing the evolution of sink 
accretion rates over the whole duration of the simulation) are based on these reruns with very high accretion rate sensitivity. We can therefore capture the infall 
rates of non-embedded PMS stars, to be compared with observational samples of measured accretion rates. 

The characteristic time-step size of our 
simulation is $\Delta t\sim2$~yr, for the highest resolution cells. Thanks to time sub-cycling (a lower resolution level can take a time step every two of the higher resolution 
level), our characteristic time-step size at the root grid resolution is  $\Delta t_{\rm root}\sim140$~yr, which also corresponds to the time step size of our sink-particle output.
We therefore have a reasonable time resolution of infall rate variations over timescales of $\sim1,000$~yr. 

In order to model ab initio the formation of individual stars, it is necessary to include a much larger scale than that of prestellar cores, to avoid imposing ad hoc 
boundary and initial conditions. By driving the turbulence on a scale of 2-4~pc, the formation of cores in our simulation is solely controlled 
by the statistics of supersonic MHD turbulence that naturally develops during the first 20 dynamical times of evolution without self-gravity. Furthermore,
a box size of $L_{\rm box}=4$~pc allows us to generate a large number of protostars, and thus to sample well the statistical distribution of the conditions of core formation
in the turbulent flow. By forming over 1,000 stars, we can use their mass distribution as an independent test to validate the simulation and give us further confidence 
of the validity of the derived infall rates. A full mass distribution is also necessary in order to correctly sample the protostellar luminosity function.

\begin{figure*}[t]
\includegraphics[width=9cm]{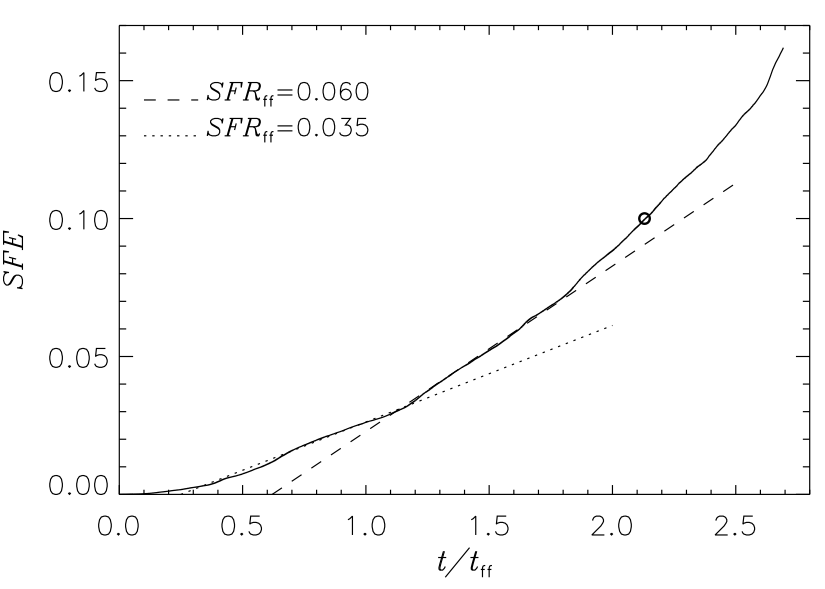}
\includegraphics[width=9cm]{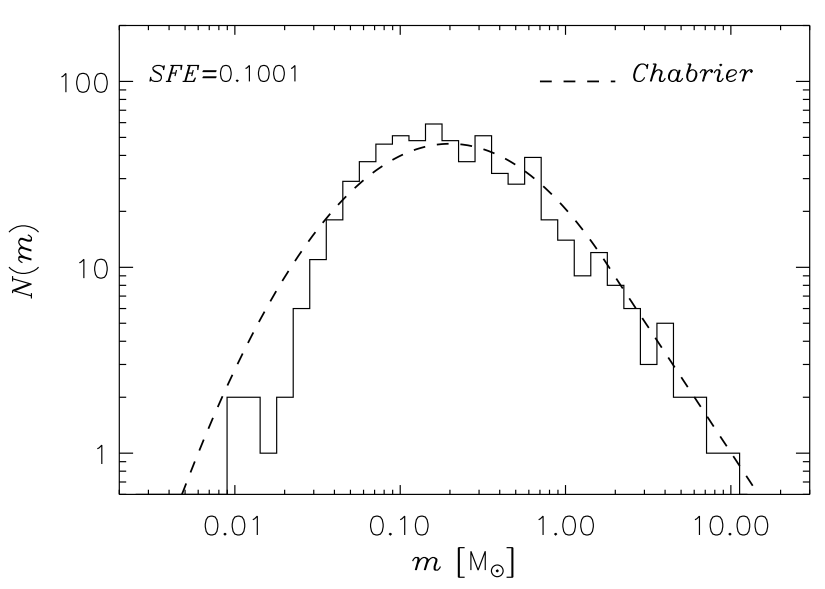}
\caption{Left Panel: $SFE$ versus time (in units of the free-fall time, and with $t=0$ corresponding to the time of formation of the first sink particle), where $SFE$ is 
defined as the total mass in sink particles divided by the initial gas mass in the simulation volume. The simulation ends at $t=3.2$~Myr, when $SFE=0.16$. The 
circle marks the time and efficiency of the snapshot used for most of the plots in the following figures. Right Panel: The mass distribution of sink particles at $t=2.6$~Myr 
after the formation of the first sink particle, when $SFE=0.10$. The dashed line shows the Chabrier IMF \citep{Chabrier2005}, connected to the Salpeter IMF 
\citep{Salpeter55} at 2~M$_{\odot}$.}
\label{sfr_imf}
\end{figure*}

The maximum spatial resolution of $\Delta x=50$~AU is partly dictated by the computational cost of the simulation and by the goal of following the evolution 
of a large number of protostars for a long time after their embedded phase. However, the main consideration in choosing the spatial resolution was 
the attempt to accurately estimate the infall rates on scales of a few 100~AU, while avoiding the complicated physics of disk formation and evolution.
The `feeding' sphere of our sink particles has a diameter of $2\, r_{\rm accr}=800$~AU, comparable to, or larger than the size of most protostellar disks.
With such values of $\Delta x$ and $r_{\rm accr}$, our sink particles are fed by infall from larger scales; disk physics is not accessible at such a resolution,
so the conversion from infall rates on the disks to accretion rates from the disks to the stars is, by design, not modeled. Although we do not compute the accretion
rate from the disks to the surface of stars, the derived infall rates will be compared to observed accretion rates, showing that infall rates are large, cannot be neglected, 
and may control both the luminosity of embedded protostars and the accretion rates of PMS stars. Furthermore, we have rerun a stretch of the simulation adding one level 
of refinement, that is increasing the spatial resolution by a factor of two, and verified that the infall rates are not significantly affected. Much deeper zoom-in simulations
\citep{Nordlund+2014} confirm that individual infall rates captured with 120 AU minimum cell size are essentially unchanged when remodeled with 2 AU minimum cell size.

Figure~\ref{maps} shows the gas density of the simulation projected along the x and y axes, and the position of all sink particles, at $t=2.6$~Myr after the formation 
of the first sink particle. The images show many young sink particles born in the densest parts of filaments, but also a large number of older ones that are no more
associated with dense gas. Some have also been ejected from binary systems, after gravitational interactions with other sinks. Due to the relatively low mean value
of $\alpha_{\rm vir}$ in this simulation, the gas has started to concentrate around a single large cloud, despite the periodic boundary conditions (the projections 
have been shifted to center the images around the densest cloud regions).

\begin{figure*}[t]
\center{
\includegraphics[width=18cm]{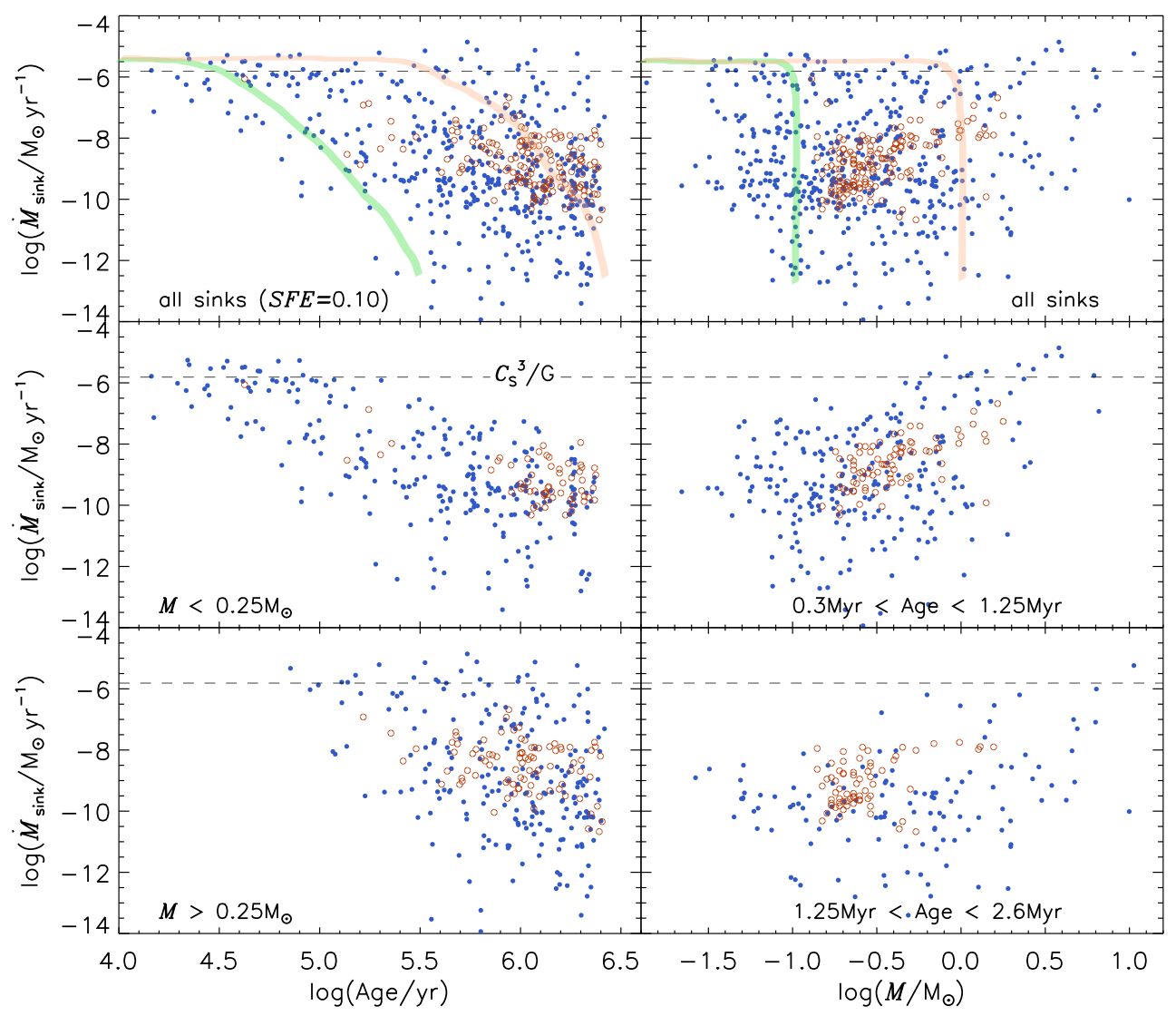}
}
\caption{Blue filled circles: Sink-particle infall rates versus sink age (left panels) and sink mass (right panels), for a snapshot of the simulation at $t=2.6$~Myr after 
the formation of the first sink particle, when $SFE=0.10$. The top panels show the 479 sink particles with detected infall rate, out of a total of 631 sink particles, 
while the middle and bottom panels show subsets for two mass intervals (left panels) and two age intervals (right panels). The thick green and pink lines in 
the top panels illustrate two idealized tracks for two sink particles with different final masses (see main text for details). Red Empty Circles: $U$--band excess
accretion rate measurements of Class II sources in the Orion Nebula Cluster by \citet{Manara+2012}. Only sources younger than 
2.6 Myr are shown.
}
\label{rates_mass_age}
\end{figure*}

\section{Star Formation Rate and Initial Mass Function}

Most of the plots in this paper are computed from a snapshot of the simulation when the $SFE=0.1$, approximately 2.6~Myr after the formation of 
the first sink particle. Although the simulation is run for 3.2 Myr from the formation of the first sink particle, when $SFE=0.16$, we have verified that our results do 
not vary with time, except for the gradual formation of the most massive sinks. We thus choose to present plots at a time when $SFE=0.1$, comparable 
to nearby star-forming regions on the scale of a few pc \citep[e.g.][]{Evans+09}. A different choice of time/SFE would not affect the conclusions of this work.

The time evolution of the star formation efficiency, $SFE=M_{\rm sink}/M_{\rm box}$, where $M_{\rm sink}$ is the total mass in sink particles, is shown 
in Figure~\ref{sfr_imf}. Up to approximately 1.1 free-fall times (corresponding to 1.34~Myr) from the formation of the first sink, the star formation rate per free-fall time,
$SFR_{\rm ff}\equiv (t_{\rm ff}/M_{\rm box}) dM_{\rm sink}/dt$, is quite low, $SFR_{\rm ff}\approx 0.035$, considering the relatively low value of $\alpha_{\rm vir}$.
It is then higher, but constant again, for almost one more $t_{\rm ff}$, with a value of $SFR_{\rm ff}\approx 0.06$. Finally, at $t\approx 1.8 \,t_{\rm ff}\approx 2.2$~Myr,
star formation accelerates, most likely because dense gas tends to accumulate around a single large cloud toward the end of the simulation,
as shown in Figure~\ref{maps}. The $SFR_{\rm ff}$ is expected to grow as star formation concentrates in regions with smaller $\alpha_{\rm vir}$ 
\citep{Padoan+2013ppvi}.

\begin{figure*}[t]
\includegraphics[width=9cm]{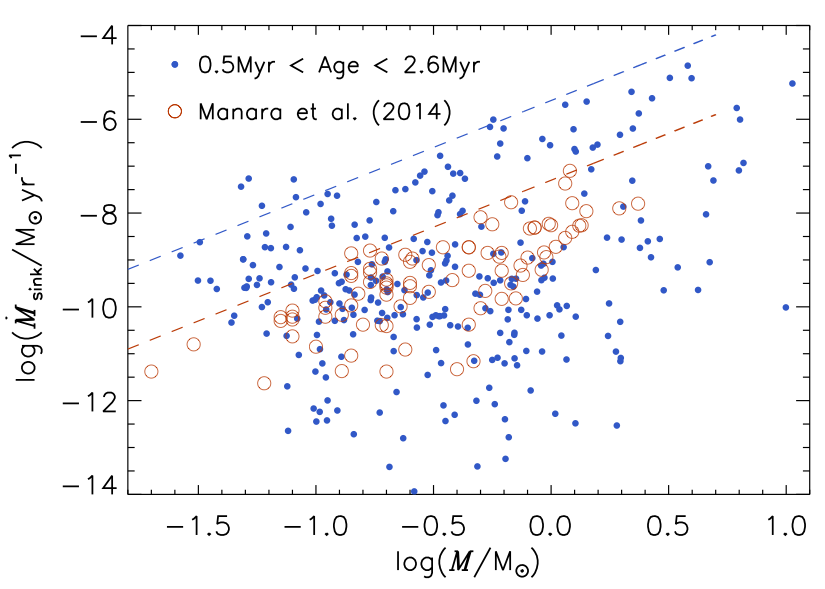}
\includegraphics[width=9cm]{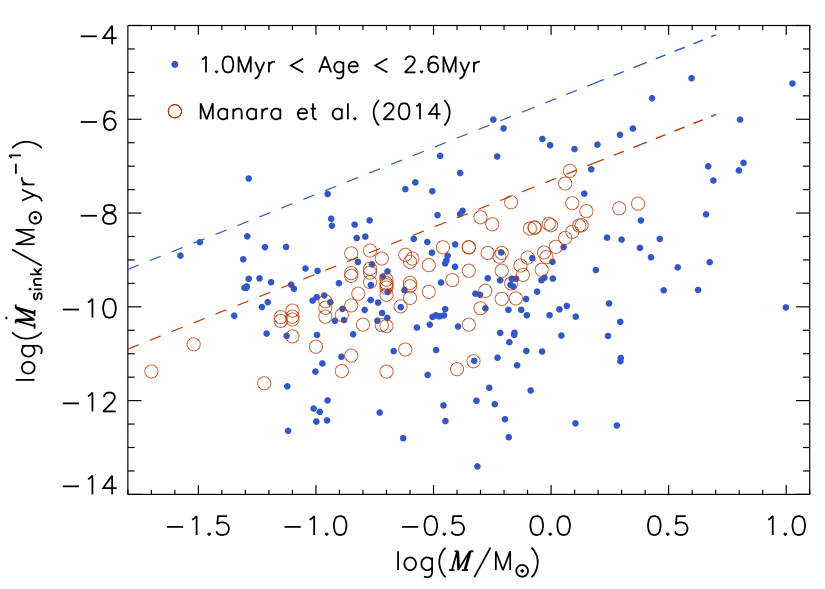}
\caption{Infall rates versus sink particle mass (blue filled circles) for sink particles older than 0.5 Myr (left panel) and 1.0 Myr (right panel). The red empty circles
show the recent compilation of accretion rates from several star-forming regions by Manara et al. (2014), mostly derived with their $UV$--excess method, based
on observations with the X-Shooter spectrograph on the VLT. The dashed lines show rates $\propto M^2$, to illustrate the approximate mass 
scaling of the upper envelopes of the plots. They are not fits to the actual values in the plots.}
\label{manara14}
\end{figure*}

We do not view the time evolution of $SFR_{\rm ff}$ in the simulation as unrealistic, or a sole consequence of not modeling stellar feedbacks. 
There is observational evidence of accelerating star formation in nearby clusters and associations \citep[e.g.][]{Palla+Stahler2000,Rygl+2013}. Although 
the evidence for age spread in star-forming regions is highly uncertain \citep[e.g.][]{Hartmann2001,Jeffries+2011,Hosokawa+2011,Preibisch2012,Soderblom+2013}, 
the lack of a significant age spread within individual clusters would also argue against a picture of constant, self-regulated star formation rate on the scale of single clusters, 
and favor a scenario where a cluster represents a local star-formation burst. This scenario is also supported by recent age determinations in massive 
star forming regions, based on a new method combining near-infrared and X-ray photometry \citep{Getman+2014}. 

The right panel of Figure~\ref{sfr_imf} shows the mass distribution of sink particles. Their mass has not been multiplied by any arbitrary efficiency factor, because 
an efficiency factor, $\epsilon_{\rm sink}=0.5$, was already adopted in the accretion model described above. This is a better approach than a global mass shift
of the sink mass distribution at the end of the simulation, because the sink mass may affect the infall rate, as it certainly does at late times, when the sink is not 
embedded in the protostellar core, and the infall rate is essentially a Bondi-Hoyle accretion. 

The mass distribution is compared with the Chabrier IMF \citep{Chabrier2005}, connected to the Salpeter IMF \citep{Salpeter55}
at 2~M$_{\odot}$. Because the simulation resolves a realistic number of binaries (down to $\sim 10$ AU separation, despite the much larger value of the exclusion radius), 
we have used the Chabrier IMF of individual stars, derived from the IMF of field 
stars, rather than that for systems. The mass distribution of sink particles follows nicely the observed IMF. Both the peak and the Salpeter slope are reproduced.
This result is stable in time, although it takes at least 1~Myr to build up the full Salpeter range, due to the relatively long timescale of formation of the more massive stars.
Based on such a comparison, our sink mass distribution may be complete down to 0.02-0.03~M$_{\odot}$. That is indeed the mass resolution limit we expect 
based on the numerical parameters of the simulation. We may therefore underestimate the total number of BDs by a factor of $\sim 2$, but this has no effect on
the conclusions of this work.

The mass distribution of sink particles is shown and compared with the observed stellar IMF because it is both an important ingredient and a fundamental constraint in 
modeling the evolution of protostars. However, a detailed discussion of the sink particle mass distribution, including a study of the effect of numerical parameters and
numerical resolution is given in a separate work (Haugb{\o}lle et al. 2014), where we demonstrate the numerical convergence of the sink particle mass distribution of 
Figure~\ref{sfr_imf}.

\section{Infall Rates and Observed Accretion Rates}

We measure the infall rate of all sink particles during the whole evolution of the simulation, that is for 3.2~Myr after the formation of the first sink particle. 
As mentioned in \S2, we can easily detect very low infall rates, of order of $\sim 10^{-11}$~\msun\ yr$^{-1}$ in the main simulation, and as low as
$\sim 10^{-14}$~\msun\ yr$^{-1}$, for the brief reruns with a lower value of $n_{\rm accr}$. One of such reruns was carried out during the time when
$SFE\approx 0.1$, thus all the following plots for $SFE=0.1$ and $t=2.6$~Myr from the formation of the first sink particle are obtained from that rerun.
All detected infall rates at $t=2.6$~Myr are shown by the plots in Figure~\ref{rates_mass_age}, both as a function of sink age (left panels) and current sink mass
(right panels). 

A striking feature of the two top plots, showing the total sample, is that their upper envelope appears to be a constant value of approximately 
$\sim 10^{-5}$~\msun\ yr$^{-1}$, independent of both age (left panel) and mass (right panel). This surprising result can be understood once 
we divide the sample into two intervals of mass (middle and bottom panels on the left) and two intervals of age (middle and bottom panels on the right).
Sink particles with masses below the peak of the mass distribution do show a decreasing infall rate as a function of age also on the upper envelope
of the plot (middle left panel). On the other hand, more massive sink particles can always be found with very high infall rates at any age (bottom left panel), 
which is the reason why they can grow to larger masses. This plot also shows that it takes a minimum of approximately 0.1~Myr to form sinks
of $0.25$~M$_{\odot}$, as almost none is found at younger ages. It is therefore already clear from these plots (and confirmed by the time evolution of individual
sinks discussed below in \S7) that the characteristic time evolution of an individual sink particle must proceed at a relatively high rate, for a certain period of time, 
and then gradually decay (although with a great variety of cases and large fluctuations around this average behavior). This picture is confirmed by the 
two plots showing the mass dependence of the infall rate for two separate age interval: once an age interval is selected, the infall rate clearly increases
with increasing sink mass. 

The other conspicuous feature of the plots in Figure~\ref{rates_mass_age} is the huge range in the values of the infall rate, at any given age or mass,
covering typically 6 orders of magnitude, except for ages below 0.1~Myr, when most sinks are in their initial phase of rapid growth. This spread can be 
partly explained by the simple picture inferred above, where individual sink particles proceed at a high constant rate for some time (the longer that time, 
the longer their final mass), and then experience a gradual decay of their infall rate. The two thick green and pink lines in the top panels of Figure~\ref{rates_mass_age} 
illustrate this simple picture by showing the idealized evolution for two sink particles ending up with two different final masses. The tracks are shown only to illustrate 
what we learn from the examination of these plots; they are not extracted from the evolution of our sink particles, nor are they numerically consistent between the left and 
right plots. As shown below (\S7), real sinks may generally follow such trends, but the fluctuations around this idealized behavior are very large, both between sinks of 
similar mass and during the evolution of individual sinks. The scatter in the plots has thus a strong contribution from large variations in sink infall rates due to 
the stochastic nature of the turbulence causing the flows converging towards the sinks.

\begin{figure*}[t]
\includegraphics[width=9cm]{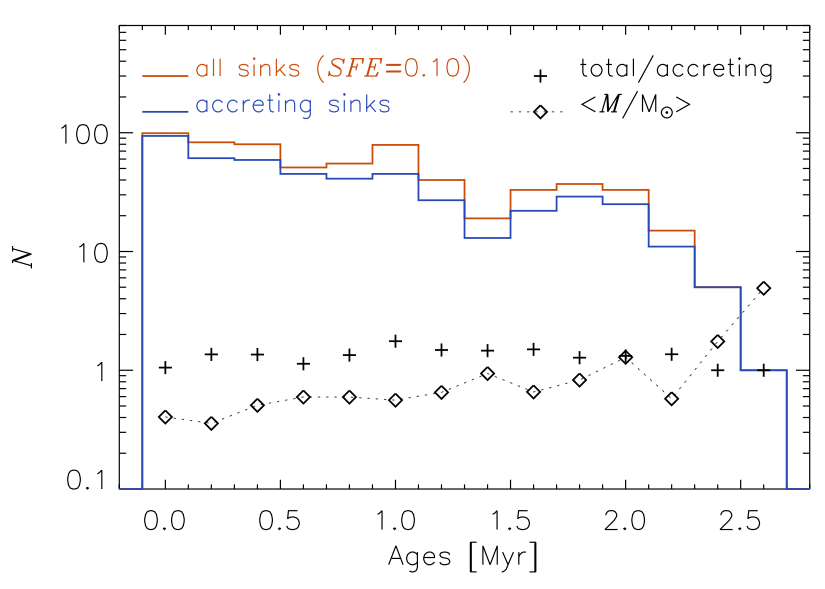}
\includegraphics[width=9cm]{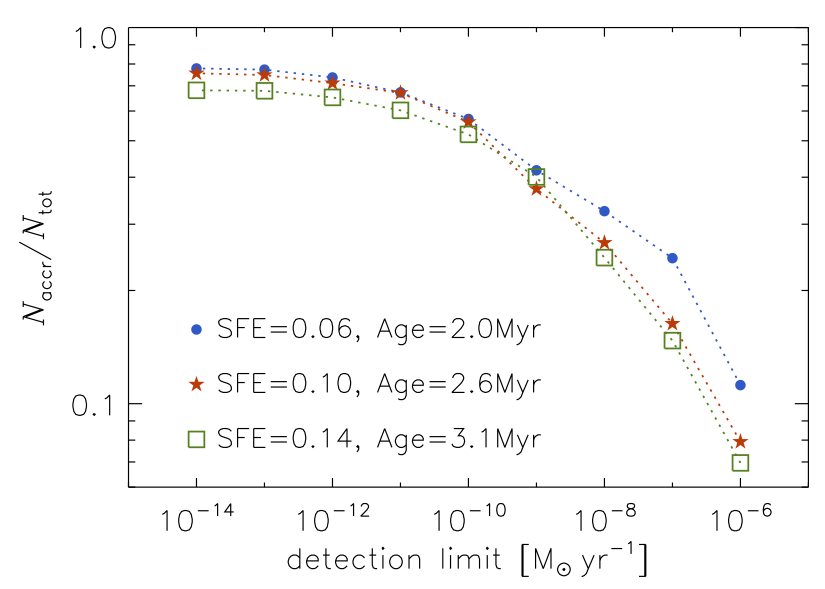}
\caption{Left Panel: Age distribution of all the sink particles at $t=2.6$ Myr (red histogram) and of the sink particles with non-zero infall rates (blue histogram). The plus
symbols show the ratio of the two histograms, and the diamond symbols the mean mass of the sink particles, in each logarithmic age interval. Right Panel: Cumulative
distribution of the infall rate for three different snapshots at $t=2.0$, 2.6 and 3.1 Myr after the formation of the first sink particle, including all sink particles (also those
with no infall rate detected.)}
\label{non_detect}
\end{figure*}

As explained in \S1, the approach of this work is to compute infall rates as a way to constrain both the growth of protostars and the accretion rates of PMS stars, 
as the infalling gas must eventually find its way to the stellar surface (except for a fraction lost in winds and jets), irrespective of the specific physical processes 
of disks that allow this to occur. While infall rates are generally not measured, determinations of accretion rates of non-embedded PMS stars have been 
carried out for several years using different observational methods. We can thus relate our predicted infall rates to the observed accretion rates, at least for sink 
particles with ages comparable to those of nearby star-forming regions where the accretion rates have been measured. In carrying out such a comparison, it should
be stressed that the infall rates in our plots correspond to the rate of growth of the sink particles. In the simulation, we use a value of $\epsilon_{\rm sink}=0.5$,
meaning that we already account for a mass loss of 50\% due to winds and jets. Our infall rates can thus be compared directly with observed accretion rates, without
any further reduction for mass loss in winds and jets (assuming $\epsilon_{\rm sink}=0.5$ is indeed a characteristic value).

The comparison with observed accretion rates \citep[e.g.][]{Muzerolle+2003,Natta+2004,Muzerolle+2005,Natta+2006,Sicilia-Aguilar+2006,GarciaLopez+2006,
Sacco+2008,Sicilia-Aguilar+2010,DeMarchi+2011,Mendigutia+2011,Ingleby+2011,Manara+2012,Rigliaco+2012,Manara+2013a,Ingleby2013,Alcala+2014} 
show that our infall rates are of the order of, or larger than the observed accretion rates. 
In Figure~\ref{rates_mass_age} we show a comparison with accretion rates from the Orion Nebula Cluster by \citet{Manara+2012}, which
is the largest observational sample to date for a single region. The observational sample includes many accretion rate measurements based on the $H\alpha$ line 
luminosity, and others based on the $U$--band excess. Because of relatively large uncertainties related to methods based on line luminosity 
\citep[][and further comments below]{Manara+2013b,Alcala+2014}, we show only the subsample based on the $U$--band excess and, in order to compare with our 
infall rates, we retain only sources with ages $< 2.6\,$Myr. Even after this selection, we are still left with a sizable sample of 173 sources. Figure~\ref{rates_mass_age} 
shows that the infall rates from our simulation have comparable values, scatter, and trends with age and mass as the observed accretion rates. In the plots where we do 
not select an age interval, the upper envelope of our infall rates is significantly higher the that of the observed accretion rates, which is to be expected because the 
observational sample does not include embedded sources.

We also have infall rate values much below the detection limit of the observed accretion rates, which is not inconsistent with the observations, once the completeness
of the survey is accounted for. According to \citet{Manara+2012}, their sample is approximately 70\% complete for Class II and III sources with masses between 0.1 
and 1 \msun . Their subset of $U$--band detections shown in Figure~\ref{rates_mass_age} is thus approximately 23\% complete. On the other hand, in the specific
snapshot of our simulation at $t=2.6\,$Myr, 76\% of the sink particles have a detected infall rate. 

\citet{Ingleby+2011} and \citet{Manara+2013b} have shown that very low values of accretion rate inferred from the luminosity of emission lines can have a significant spurious 
component due to chromospheric activity, and are thus highly uncertain. \citet{Manara+2013b} have developed a more precise method to determine 
accretion rates based on the $UV$ excess, and have carried out an observational campaign of several star-forming regions with 
the VLT X-Shooter spectrograph \citep[][Manara et al. 2014]{Rigliaco+2012,Alcala+2014}. In Figure~\ref{manara14}, we plot their measurements (only the detections)  
together with our predicted infall rates at $t=2.6$~Myr from the formation of the first sink particle. Because the observations include only non-embedded Class II sources, 
we have only plotted infall rates of sink particles older than 0.5~Myr (left panel of Figure~\ref{manara14}), the estimated approximate 
duration of the Class I phase, according to \citet{Evans+09}. 

Because of the very uncertain relation between age and protostellar class, in the right panel of Figure~\ref{manara14} 
we also show the comparison including only sink particles older than 1~Myr. In both cases, our predicted infall rates are comparable to, or larger than the 
observed accretion rates. Some of our infall rates are lower than the detection limit of the observations, which may still be consistent with the observations, given that
this sample is far from complete. Both the simulated infall rates and the observed accretion rates show a well defined upper envelope, scaling approximately as $M^2$ 
(dashed lines). Even in the case of the older sinks (right panel of Figure~\ref{manara14}), the upper envelope of the infall rates from the simulation is significantly above 
the upper envelope of the observed stellar accretion rates. That is because, even after this age selection, the sink particles with the highest accretion rates are somewhat embedded 
and would appear as Class 0 or I, and thus would not be part of the observational sample.

\begin{figure*}[t]
\includegraphics[width=9cm]{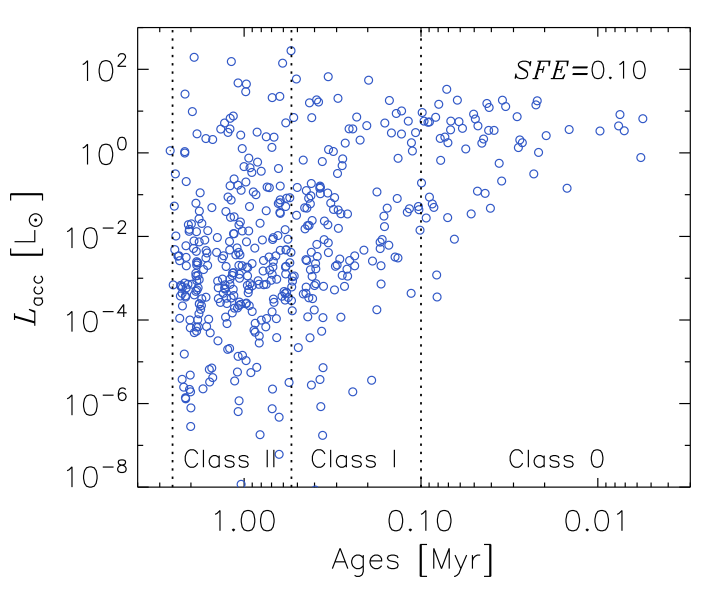}
\includegraphics[width=9cm]{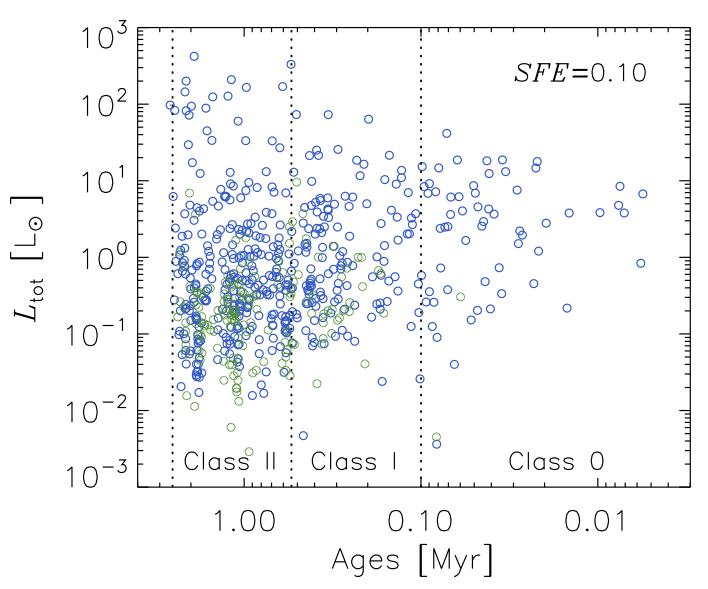}
\caption{Left Panel: Accretion luminosity versus sink particle age computed from the snapshot of the simulation at $t=2.6\,$Myr. The vertical dotted lines separate the
different classes, based on the class age estimates by \citet{Evans+09}. Right Panel: Total protostellar luminosity versus sink particle age for the same
snapshot of the simulation as on the left panel. The blue circles are for the same sink particles as in the left panel, while the green circles are for sink particles with no 
infall, thus $L_{\rm acc}=0$, not shown on the left panel.}
\label{Lacc}
\end{figure*}

This comparison with the observations suggests that infall rates can be important even after the embedded protostellar 
phase, for PMS ages of approximately 1-3 Myr. A more rigorous comparison of the simulation with the observations requires radiative transfer 
calculations, in order to generate synthetic observations to establish a reliable association of our sink particles with PMS classes (Frimann et al. 2014). 
This will be pursued in a separate work.
The completeness of the observational samples should also be accurately estimated. As far as the simulation is concerned, the fraction of sink particles with a detected infall 
rate is always very large (larger than 60\%) for all sink ages, as shown in the left panel of Figure~\ref{non_detect}. In the right panel of Figure~\ref{non_detect}, we
also show the cumulative distributions of the infall rate for three different snapshots, corresponding to $t=2.0$, 2.6 and 3.1 Myr from the formation of the first sink
particle. The fraction of accreting to non-accreting sink particles does not grow significantly with decreasing infall rate for infall rates below approximately 
$10^{-11}$~\msun\ yr$^{-1}$, and does not depend strongly on the age of the star-forming region, at least in the range between 2.0 and 3.1 Myr. Current surveys 
can reach a detection limit of $10^{-11}-10^{-12}$~\msun\ yr$^{-1}$ for low mass PMS stars. Thus, based on the cumulative distributions of Figure~\ref{non_detect},
such surveys should yield detections for half or more of the sources, at least in star-forming regions not much older than 3~Myr.

\section{The Luminosity Problem}

Accretion rates are not directly measured in deeply embedded protostars. They are inferred from their luminosity, because in very young protostars,
say less than 0.1~Myr, the accretion luminosity is much larger than the protostellar luminosity. With reasonable assumptions about the protostellar 
radius and mass, the accretion rate is then approximately given by assuming that most of the gravitational energy of the accreting gas is released as radiation,
$L_{\rm acc} \approx a \,M \dot{M}/r$, where $M$ and $r$ are the protostellar mass and radius respectively, $\dot{M}$ is the accretion rate, and $0<a<1$ is an 
efficiency factor that depends on details of the accretion from the disk and envelope to the stellar surface. If one then infers the duration of the embedded protostellar 
phase from the relative number of embedded protostars and T Tauri stars, and from the estimated lifetime of T Tauri stars, one gets an average accretion rate that 
yields an accretion luminosity an order of magnitude larger than the characteristic luminosity of embedded protostars (assuming $a\approx 1$). This is known as the 
`luminosity problem', first discovered by Kenyon et al. (1990), who also proposed several solutions. A more theoretical 
view of the same problem, also first recognized by Kenyon at al. (1990), is that the accretion rate due to gravitational collapse cannot be smaller than 
approximately $c_{\rm s}^3/G$ (Stahler, Shu, and Taam 1980), which is $\approx 10^{-6}$~\msun\ yr$^{-1}$  for a characteristic molecular cloud 
temperature of 10~K, approximately 10 times larger than the average accretion rate inferred from observations.

\begin{figure*}[t]
\includegraphics[width=9cm]{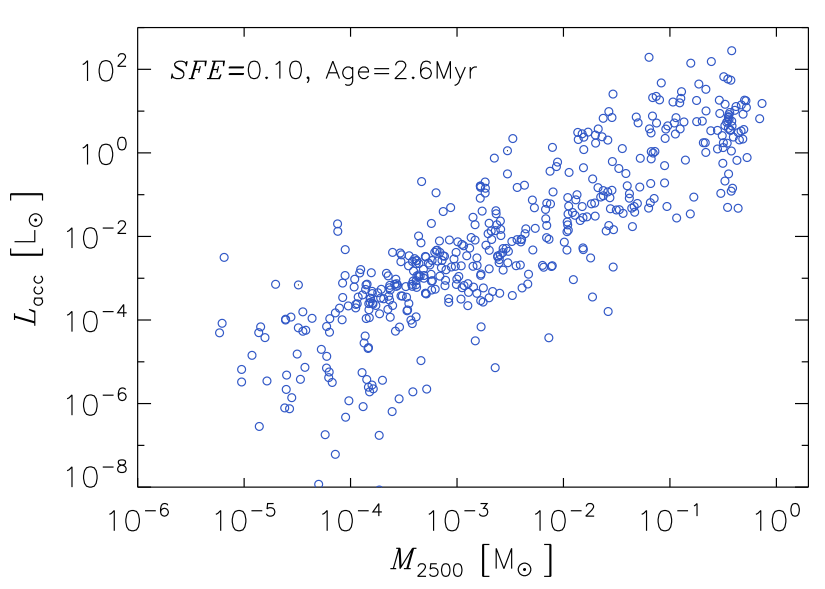}
\includegraphics[width=9cm]{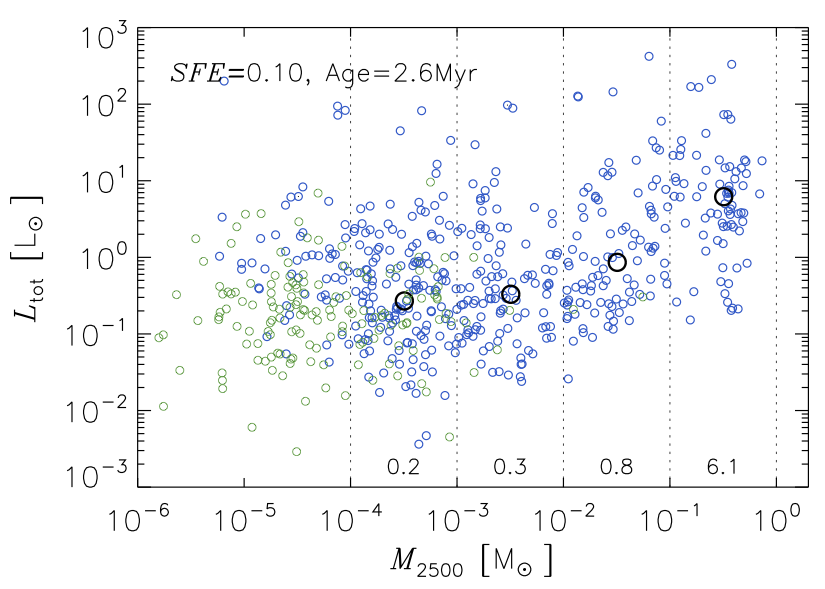}
\caption{Left Panel: Accretion luminosity versus envelope mass within 2,500 AU from a sink particle, for the simulation snapshot at $t=2.6\,$Myr. 
Right Panel: Total protostellar luminosity versus envelope mass. As in Figure~\ref{Lacc}, the green circles show sink particles with no infall. The four,
larger, black circles show the median values of the total protostellar luminosity within the logarithmic age intervals marked by the dotted vertical lines.
Those values are also given by the numbers at the bottom of each interval. The median luminosity decreases with decreasing envelope mass. 
}
\label{lum_m25}
\end{figure*}

As already suggested in Kenyon et al. (1990) and extensively investigated in more recent studies 
\citep[e.g.][]{Dunham+2010,Offner+McKee2011,Dunham+Vorobyov2012,Myers2012,Dunham+2013}, the luminosity problem
puts strong constraints on the time evolution of the accretion rate of protostars. We can thus test the validity of our predicted 
protostellar infall rates by computing the resulting protostellar luminosities and comparing them with the observed values. In doing this,
we assume that the accretion rate is the same as the infall rate, because the infalling gas must eventually find its way onto the stellar surface,
and the infalling mass cannot reside on the protostellar disk for a very long time, because disk masses would then be much larger than indicated
by observations. Our infall rates already account for a 50\% mass loss by winds and jets ($\epsilon_{\rm sink}=0.5$ in the simulation, for both stellar masses and
accretion rates). Assuming that the accretion rate is equal to the infall rate, and with a choice of protostellar radii, we can then compute the accretion luminosity.
We neglect energy losses related to winds and jets, so our predicted luminosity is an upper limit to the accretion luminosity. We also neglect disk instabilities 
that may cause variations of the accretion rates on very short timescales, so the amplitude and frequency of our predicted time variations of the
accretion rate are probably underestimated. Both consequences of our approximations go in the direction of hindering a possible solution to the luminosity 
problem, so they do not ease the luminosity constraint on the infall rates. 

The radius of a young, low mass, accreting protostar depends on several factors such as initial conditions (for example the initial radius adopted in the stellar
evolution calculations), the time evolution of its accretion rate, and the fraction, $\alpha$, of the internal energy of the accreting material that is absorbed by the protostar
(e.g. Hartmann et al. 1997, Baraffe et al. 2009). The computation of the stellar radius as a function of time for protostars corresponding to our individual sink particles is 
beyond the scope of this work. We simply assume that the protostellar radius is given by $r=3 R_{\odot} (M/M_{\odot})^{0.5}$, which gives a reasonable 
approximation to values derived in Hartmann et al. (1997) for different cases with accretion rates of $\sim 10^{-6}-10^{-5}$~\msun\ yr$^{-1}$ (as in our sink
particles at young ages), and for the `cold accretion' case of $\alpha \ll 1$. The evolution of the stellar radius after accretion has subsided is not important, because
the stellar luminosity will then be much larger than the accretion luminosity. With such a choice of stellar radius, we introduce an uncertainty in the total
luminosity (stellar plus accretion luminosity) of at most a factor of two on the average. The accretion luminosity is then given by 
$L_{\rm acc}= \epsilon(1-\alpha)M \dot{M}/r$, where $\epsilon$ is the ratio of internal to gravitational energies of the accreting material. 
Although $\epsilon<1$, with the precise value depending on the details of the accretion process (Hartmann et al. 1997), for simplicity, we choose $\epsilon=1$,
as well as $\alpha=0$. 

All our assumptions (infall rate equal to accretion rate, $\alpha=0$, and $\epsilon=1$) lead to an overestimate of the accretion luminosity, and they are thus conservative
from the point of view of solving the luminosity problem. The value of $L_{\rm acc}$ computed for all our sink particles at $t=2.6$~Myr after the formation of the fist sink
are plotted in the left panel of Figure~\ref{Lacc} (except for the cases with no detected infall, or $L_{\rm acc}=0$). The plot shows a large scatter in $L_{\rm acc}$, 
increasing with sink age, from approximately three orders of magnitude before 0.1~Myr, to 8 or more orders of magnitude after 0.1~Myr. The upper envelope of the 
plot grows slightly with age, by approximately one order of magnitude, as it corresponds to the highest accretion rates that, at large ages, is found on the most massive
stars ($L_{\rm acc}$ scales linearly with $M$). 

\begin{figure*}[t]
\includegraphics[width=9cm]{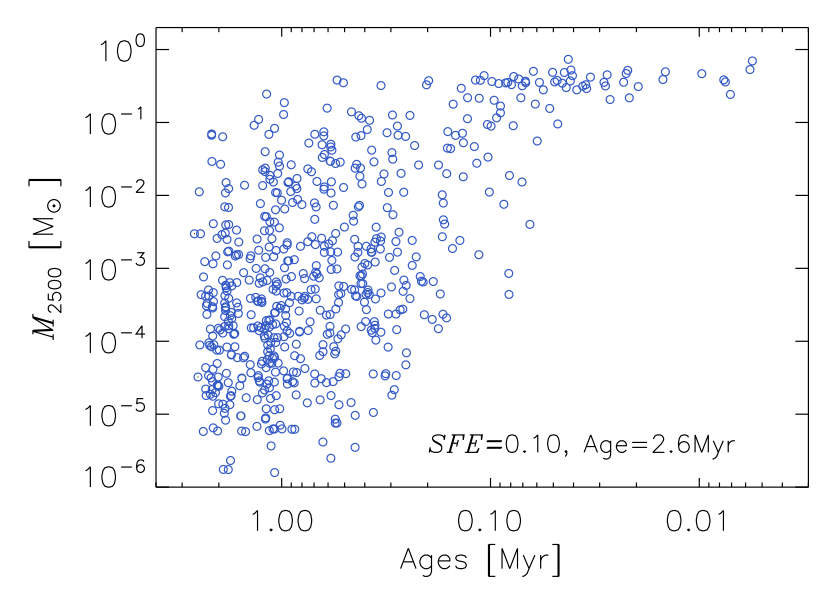}
\includegraphics[width=9cm]{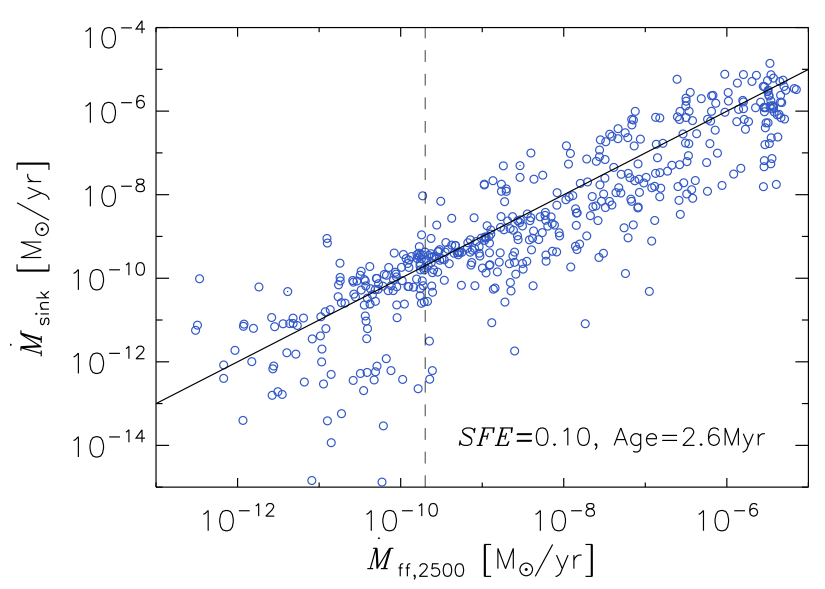}
\caption{Left Panel: Envelope mass versus sink particle age, for the simulation snapshot at $t=2.6\,$Myr. The upper envelope of this scatter plot is nearly independent
of sink age, because relatively massive stars can preserve a rather large envelope mass (and a large infall rate) for 1 Myr or longer. Right Panel: Infall rate of sink particles
measured in the simulation versus infall rate computed as ${\dot M}_{\rm ff,2500}=\epsilon_{\rm sink}\,M_{2500}/t_{\rm ff,2500}$, where $t_{\rm ff,2500}$ is the 
free-fall time corresponding to the average gas density within 2,500 AU from the sink particles. The solid line marks the equality between the two infall rates, 
${\dot M}_{\rm sink}={\dot M}_{\rm ff,2500}$, while the dashed vertical line shows the value of ${\dot M}_{\rm ff,2500}$ corresponding to the smallest envelope 
masses in the HOPS sample (Fisher et al. 2014)}
\label{m25}
\end{figure*}

The classification of young stars is based on the bolometric temperature, that is on their spectral energy distribution. It is believed to be generally related to protostellar age,
but with large uncertainties \citep[e.g.][]{Dunham+2010}, and a one-to-one relation between class type and age is not possible for individual protostars. In Figure~\ref{Lacc}, 
we mark age boundaries for the different classes (vertical dashed lines) based on the duration of each class type derived by \citet{Evans+09}. Because radiative transfer 
calculation to establish the class type of each of our sink particle is beyond the scope of this work, we simply rely on this relation between class type and age in order to 
compare with observed protostellar luminosities. However, we have verified that all our sinks classified as Class 0 in Figure~\ref{Lacc} are indeed deeply embedded, 
so those with low accretion luminosity have either low mass, or low infall rate (despite their significant envelope mass), or both. 

In order to compare with the observations, we express the total luminosity, $L_{\rm tot}$, as the sum of the accretion luminosity, the stellar luminosity, $L_{\rm star}$, 
and the envelope luminosity, $L_{\rm ISRF}$, $L_{\rm tot}=L_{\rm star}+L_{\rm acc}+L_{\rm ISRF}$, as in \citet{Young+Evans2005}. 

The stellar luminosity is taken 
from the evolutionary tracks of \citet{DAntona+Mazzitelli97,DAntona+Mazzitelli98}, using the mass and the age of the sink particle, and adding 0.1 Myr to the tabulated 
ages before performing the interpolation to compute the stellar luminosity, as in \citet{Young+Evans2005}, to account for the approximate time before the start of 
deuterium burning \citep{Stahler83}. The uncertainty introduced by this procedure is not important, because, in the first few 0.1 Myr, $L_{\rm star}$
is usually much smaller than $L_{\rm acc}$. 

The envelope luminosity, due to the thermal emission of dust grains, is computed as
$L_{\rm ISRF}=0.6 \,{\rm L}_{\odot}\,M_{\rm env}/{\rm M}_{\odot}$, where $M_{\rm env}$ is the envelope mass, which corresponds approximately to the emission
of silicate grains with size $a=0.1 \,\mu$m, temperature $T=10\,$K, and Planck--averaged emissivity 
$\langle Q_{\rm abs}\rangle=1.4\times10^{-3} (a/1\,\mu{\rm m})(T/10\, {\rm K})^2$ \citep{Draine+Lee84}. 
We compute $M_{\rm env}$ as the mass within a sphere with a radius of 5,000~AU centered on the sink particle, and when multiple sinks share the same envelope,
simply divide the envelope mass by the number of sinks. 

This evaluation of $L_{\rm ISRF}$ is quite uncertain. It will be improved, in a separate work, with radiative 
transfer calculations to try to mimic the way in which the envelope luminosity truly enters the bolometric luminosity derived from observations of embedded protostars.
The radiative transfer calculations should also account for the local enhancement of the interstellar radiation field due to the massive sink particles in the simulation, 
which could significantly enhance the dust temperature of at least some of the envelopes (see Frimann et al. 2014). The uncertainty in $L_{\rm ISRF}$ affects only the 
total luminosity of a few sink particles of very young age (large envelope mass and low stellar luminosity) and very low infall rate (low accretion luminosity). 
In other words, $L_{\rm ISRF}$ only determines the lower envelope of the scatter plot of total luminosity versus age for very young protostars (up to 0.1-0.2 Myr of age), 
and the low--luminosity tail of the protostellar luminosity function of embedded protostars.

The total luminosity, $L_{\rm tot}$, is plotted versus the sink age in the right panel of Figure~\ref{Lacc}. Apart from the large uncertainty in the conversion between age 
and protostellar class mentioned above, this plot could be directly compared with Figure~13 in \citet{Evans+09}, which illustrates the luminosity problem in the context 
of the c2d Spitzer Legacy project \citep{Evans+03}. The observations show a scatter in $L_{\rm tot}$ of three orders of magnitude, between 0.1 and 100 L$_{\odot}$, 
for Class 0 and Class I protostars, which cannot be explained by standard collapse models.  The right panel of our Figure~\ref{Lacc} shows approximately the same 
range of values of $L_{\rm tot}$. We can therefore conclude that, for realistic infall--rate (accretion--rate) histories of protostars, as obtained from our simulation, there 
is no luminosity problem. Observed protostellar luminosities are in agreement with our simulation. The uncertainties introduced by our approximations 
would go in the direction of over-estimating the luminosity, and so they cannot be the reason why our simulation does not yield too high luminosities. 

Besides the low average luminosity, the major challenge presented by the observations is to explain the large luminosity scatter of three orders of magnitude. 
Such a scatter is successfully reproduced in our simulation as a direct consequence of the large scatter in infall rates, which is a robust result irrespective of the various
approximations adopted to estimate the total luminosities. The large scatter in infall rates is to be 
expected in star-forming regions, where the infall is ultimately controlled by stochastic converging flows in the turbulent ISM. 

\begin{figure*}[t]
\includegraphics[width=9cm]{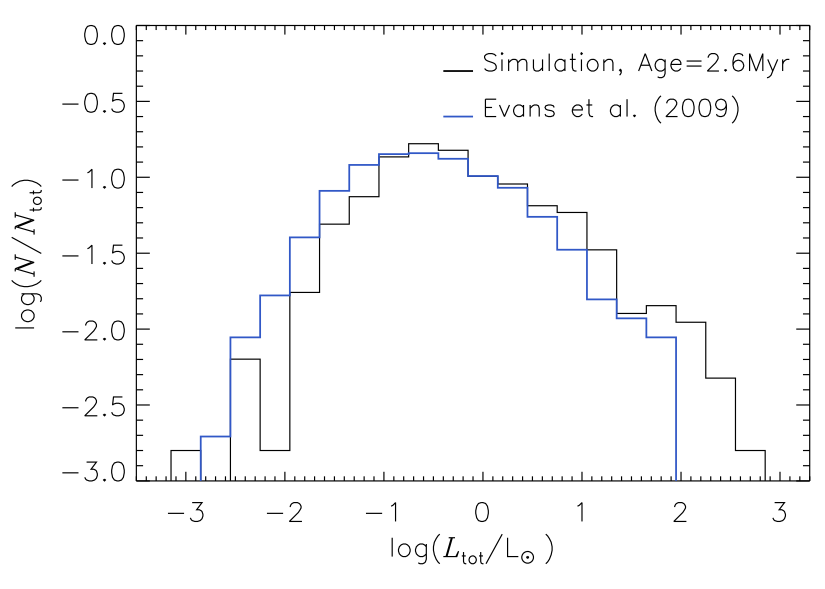}
\includegraphics[width=9cm]{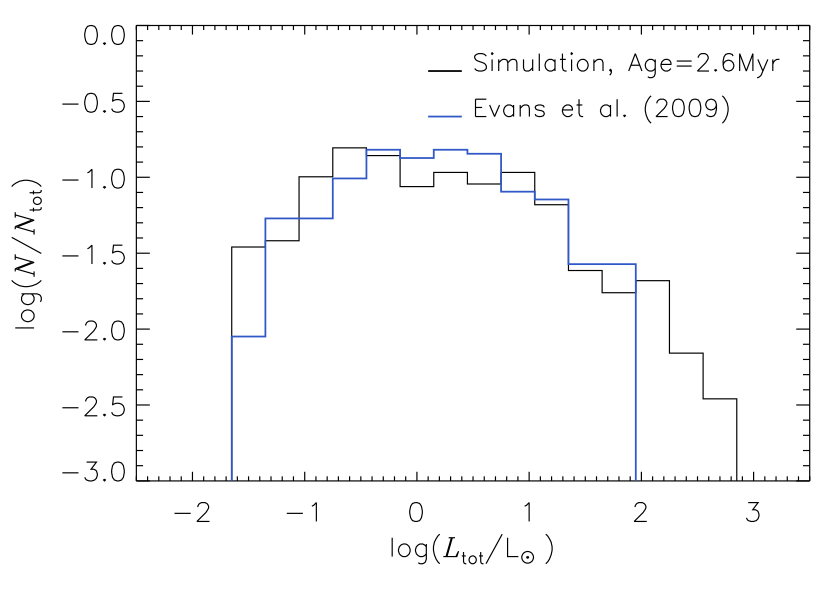}
\caption{Left Panel: Luminosity function using all the 631 sink particles found in the simulation at $t=2.6$~Myr (black histogram), and for the full sample of 1,024 sources
from Class 0 to Class III from \citet{Evans+09} (blue histogram). Right Panel: Comparison of the PLF of sink particles with $M_{2500}> 0.001$~\msun (black histogram), 
with the PLF for the subset of 122 embedded protostars identified in \citet{Evans+09} and used in the PLF of \citet{Dunham+2010} (blue histogram). In all four PLFs,
the number of protostars in each luminosity bin is normalized to the total number of protostars in the sample.}
\label{LF}
\end{figure*}

Given the uncertain link between protostellar class and age, and the difficulty to assign a protostellar class to a sink particle based on the simulation data, even
if radiative transfer calculations had been carried out (e.g. the spectral energy distribution could be sensitive to small-scale protostellar disk structure that is not 
captured in the simulation -- see Frimann et al. 2014), a more direct way to compare our results with the observations is to use the envelope mass. Upcoming 
protostar surveys, such as the Herschel Orion Protostar Survey (HOPS) \citep{Manoj+2013,Fisher+2013,Stutz+2013}, will be suitable for such a comparison of
envelope masses, as an alternative to bolometric temperature. 

Rather than trying to define the total envelope mass, Fischer et al. (2014) 
have computed the mass within 2,500~AU from each protostar, $M_{2500}$, and plotted the total luminosity as a function of $M_{2500}$. In the left panel of
Figure~\ref{lum_m25}, we show the accretion luminosity as a function of $M_{2500}$ computed from our simulation at $t=2.6$~Myr; in the right panel of the same
figure, we plot the total luminosity. The comparison of the two plots shows that $L_{\rm acc}$ is the main contribution to $L_{\rm tot}$ for protostars with the largest 
values of $M_{2500}$. The right-hand side panel of Figure~\ref{lum_m25} can be directly compared to the corresponding plot in Fischer et al. (2014). The main 
difference is that we can `detect' much smaller values of $M_{2500}$ ($\sim 10^{-6}$~M$_{\odot}$) than in the HOPS data ($\approx3\times10^{-4}$~M$_{\odot}$). 
As in the observations, we find an overall decrease in luminosity with decreasing envelope mass, at least in the range $10^{-3}$~M$_{\odot}< M_{2500}< 1$~M$_{\odot}$.
We also find very similar scatter and mean values of $L_{\rm tot}$ as in the observations, showing again that there is no protostellar luminosity problem when we account 
for realistic infall rates as those given by our simulation. 

To provide further insight into the role of $M_{2500}$, we also plot, in the left panel of Figure~\ref{m25}, $M_{2500}$ versus the sink particle ages. One can see that,
while the smallest values of $M_{2500}$ decrease rapidly with increasing ages, the upper envelope of this scatter plot is nearly independent of age.  
This is because the formation of stars in the Salpeter range of masses requires an extended period of time with a relatively large infall rate, the longer that
time, the more massive the final star (on the average). 

Observed values of $M_{2500}$ may be used for an approximate estimate of the envelope infall rate, by dividing the envelope mass by its free--fall time.
In the right panel of Figure~\ref{m25}, we compare such infall rate estimates with the infall rate of the sink particles measured in the
simulation. We have defined the estimated envelope infall rate as ${\dot M}_{\rm ff,2500}=\epsilon_{\rm sink}\,M_{2500}/t_{\rm ff,2500}$, because the sink particle
infall rate already accounts for the efficiency factor, $\epsilon_{\rm sink}$. The solid line marks the equality between the two infall rates, 
${\dot M}_{\rm sink}={\dot M}_{\rm ff,2500}$, while the dashed vertical line shows the value of ${\dot M}_{\rm ff,2500}$ corresponding to the smallest envelope 
masses in the HOPS sample (Fisher et al. 2014). The two infall rates are linearly correlated, but with a very large total scatter, of roughly 3 orders of magnitude.

\section{The Protostellar Luminosity Function}

Besides scatter plots of bolometric luminosity versus bolometric temperature, versus class type, versus age (right panel of Figure~\ref{Lacc}) 
or versus envelope mass (right panel of Figure~\ref{lum_m25}), the luminosity problem can be addressed through the protostellar luminosity function (PLF) of
embedded protostars (see \citet{Dunham+2014} for a recent review). The PLF depends on the mass distribution of protostars, which yields the stellar IMF,
and on the total protostellar luminosity (the sum of the luminosities of the star, the envelope, and the accretion shock, for embedded protostars --see the previous section). 

\begin{figure*}[t]
\includegraphics[width=9cm]{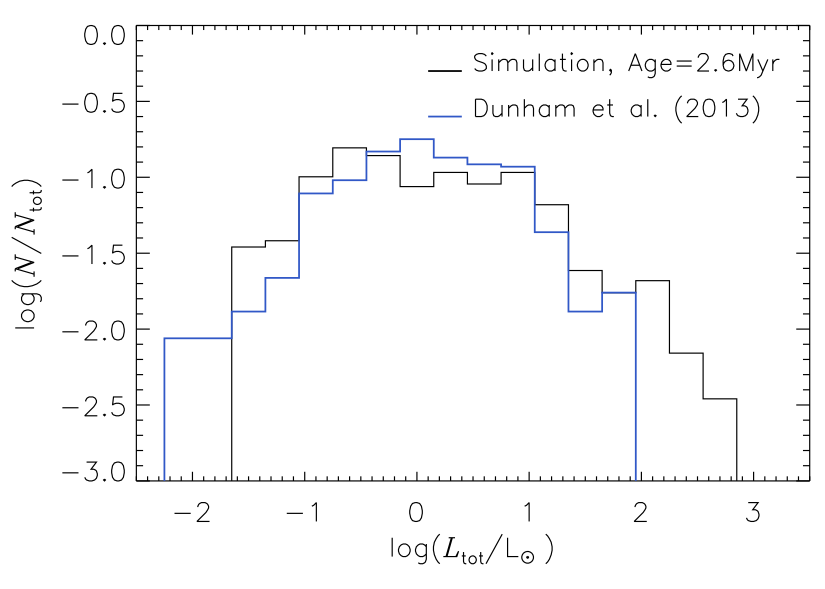}
\includegraphics[width=9cm]{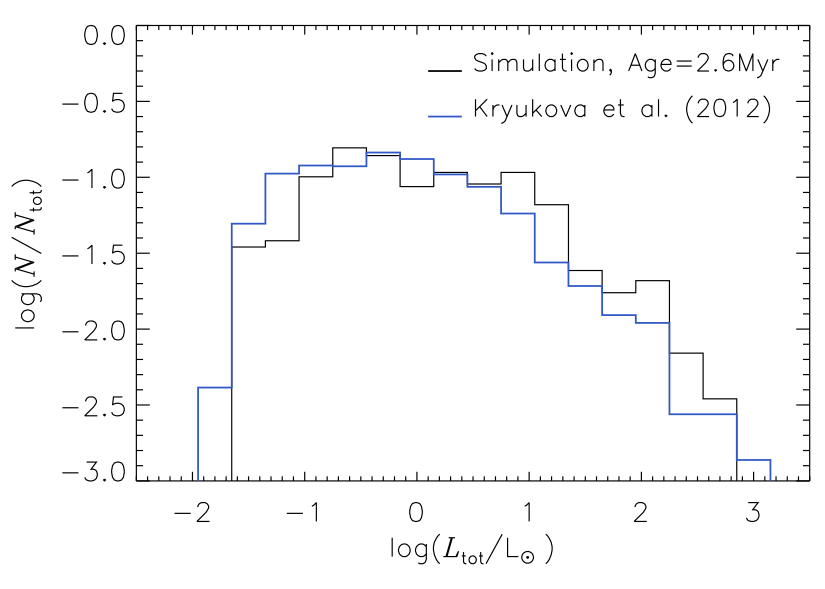}
\caption{Left Panel: Comparison of the same embedded sink-particle PLF as in the right panel of Figure~\ref{LF} (black histogram), with the extended PLF
of 230 protostars by \citet{Dunham+2013} (blue histogram), lacking the sub-mm luminosity for 100 of the protostars. Right Panel: Same sink-particle PLF
as in the left panel (black histogram), compared with the PLF of all 728 protostars from \citet{Kryukova+2012}. All PLFs are normalized as those in Figure~\ref{LF}.}
\label{PLF}
\end{figure*}

Before considering the PLF of embedded protostars, we compare, in the left panel of Figure~\ref{LF}, the luminosity function (LF) of all the 631 sink particles found in the simulation at 
$t=2.6$~Myr (including those with very low envelope mass), with the observed LF of the full sample of \citet{Evans+09}, composed of 1,024 sources,
from Class 0 to Class III. This comparison is useful to bring out differences between the IMF of the simulation and that of the observational sample. 
At $t=2.6$~Myr, the simulation has 28 sink particles more massive that 2~\msun\ and 11 more massive than 4~\msun\  (the largest mass is 
10.6~\msun). This is most likely a larger number of relatively massive stars than in the c2d sample, probably because the mean density in the simulation  
box, and in the main cluster-forming regions of the simulation are more typical of a region like Orion than of those in the c2d survey. The IMF of the simulation 
is also somewhat incomplete at small BD masses, due to the limited numerical resolution (see Haugb{\o}lle et al. 2014 for details about the numerical 
convergence of the IMF of this simulation). The left panel of Figure~\ref{LF} shows some evidence for both an overabundance of massive stars, with 
luminosity in excess of 100~L$_{\odot}$, and a shortage of BDs, with luminosity less than 0.01~L$_{\odot}$, in the simulation relative to the observations. 
On the other hand, the peak of the observed LF and its overall shape between 0.01 and 100~L$_{\odot}$ are very nicely matched by the simulation.  

While the low-luminosity tail of the LF is uncertain, due to our simplified estimate of the envelope luminosity (see the previous section) and to uncertainties in the BD stellar
evolutionary tracks used to compute the stellar luminosities, the excess of massive stars is a robust result that is confirmed by the comparison
of the PLFs of embedded protostars, shown in the right panel of Figure~\ref{LF}. In this plot, we have used the 112 sources with detected envelopes 
from \citet{Evans+09}, that is the observed PLF in \citet{Dunham+2010}, and the 288 sink particles with $M_{2500}> 0.001$~\msun (corresponding 
to envelope masses within 5,000~AU larger than approximately 0.01~\msun, and total envelope masses typically a few times larger than that). 
Besides the existence of a few objects more luminous than $100$~L$_{\odot}$, the PLF of embedded sink particles
matches nicely the observed embedded PLF, showing again that the simulation does not imply any luminosity problem. 

All the embedded protostars in the PLF of \citet{Dunham+2010} have bolometric luminosity estimated with the aid of sub-mm measurements.   
\citet{Dunham+2013} extended the sample from 112 to 230 protostars, but 100 of the new protostars lack sub-mm measurements, so their 
luminosity is underestimated by a factor of approximately 2.6, on the average, and, in some cases, up to a factor of 10 \citep{Dunham+2013}.
The low-luminosity tail of this PLF is thus rather uncertain. As can be seen in the left panel of Figure~\ref{PLF}, this PLF has an excess of low
luminosity protostars, relative to the PLF of \citet{Dunham+2010}, almost certainly caused by the lack of sub-mm measurements. Apart from that
feature, this PLF is also nicely matched by the embedded PLF of our sink particles. 

\citet{Kryukova+2012} derived a PLF from a very large sample of 728 protostars, from both low-mass and high-mass star-forming regions, 
also based on Spitzer data. None of their protostellar candidates has sum-mm measurements, but they apply a correction to convert the
mid-IR luminosity to bolometric luminosity, based on the spectral-energy-distribution slope, which they calibrate on the c2d protostars with 
well-determined bolometric luminosity. They present the PLF for each of the nine regions they observe, besides the PLFs obtained by combining
all low-mass star-forming regions, and all high-mass ones, in order to study the role of environment in shaping the PLF. We postpone the discussion
of environment to the next section, and, in the right panel of Figure~\ref{PLF}, we instead compare the PLF of all 728 protostars from 
\citet{Kryukova+2012} with the PLF of our embedded sink particles. Because this observational sample contains also high-mass star-forming regions,
the PLF reaches higher total luminosities than in the samples by \citet{Evans+09} and \citet{Dunham+2013}, and so it compares even better
with our simulation results than the two previous PLFs. 

\begin{figure*}[t]
\includegraphics[width=18cm]{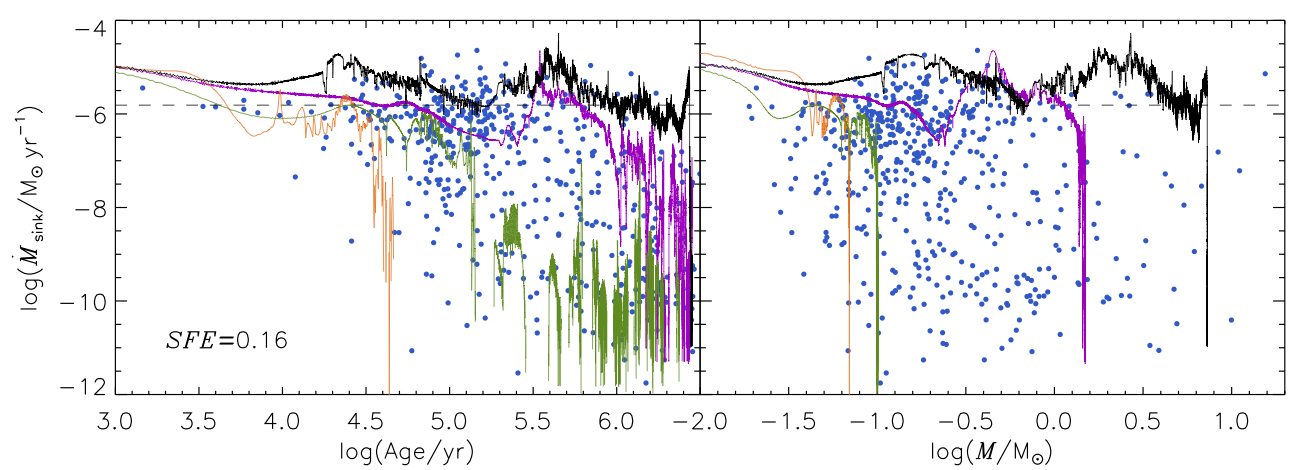}
\caption{Evolution of the infall rate of four characteristic sink particles ending up with different final masses, as a function of sink age (left panel) and sink mass (right panel).
The blue filled circles show the values of infall rate, age, and mass of all the sinks in the least snapshot of the simulation, at $t=3.2$ Myr, when $SFE=0.16$. The observed
scatter is clearly accounted for both by the evolution of a single sink, and by differences between the tracks of different sinks.}
\label{tracks_age_mass}
\end{figure*}
\begin{figure}[t]
\includegraphics[width=8.7cm]{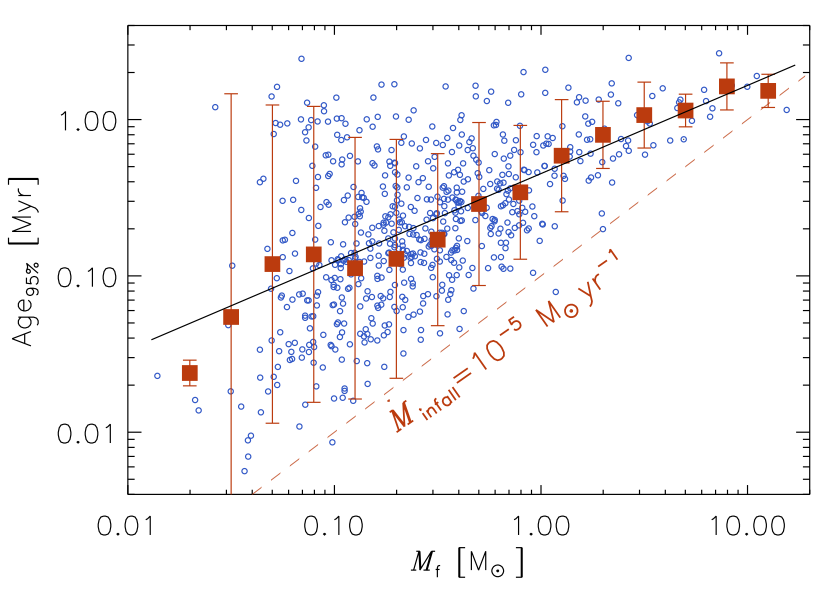}
\caption{Age of sink particles when 95\% of their final mass has been assembled, versus sink particle final mass, defined as the sink-particle mass at the end of
the simulation, at $t=3.2$ Myr. The empty circles show all the sinks found at $t=2.6$ Myr. Most of these sinks have indeed stopped growing significantly by the 
end of the simulation. The filled squares show the average Age$_{95\%}$ computed inside logarithmic intervals of the final mass. The solid black line is a linear fit
to the logarithmic values of Age$_{95\%}$ versus final mass, giving Age$_{95\%}=0.45\,{\rm Myr} \,(M_{\rm f}/{\rm M}_{\odot})^{0.56}$.}
\label{age95}
\end{figure}

The match between our sink-particle PLF and the PLF of \citet{Kryukova+2012} is excellent over the whole luminosity range, covering five orders 
of magnitude. The slight lack of intermediate and high-luminosity sources in the PLF of \citet{Kryukova+2012} may be due to the fact that protostars 
from saturated areas of the 24 $\mu$m maps of Orion were not included in the survey. Furthermore, the slight excess of protostars with luminosity
below 0.1 L$_{\odot}$ would be eliminated if we included a correction to remove the residual contamination by background galaxies, and edge-on  
and reddened Class II sources, as in \citet{Kryukova+2012}. We do not attempt such a correction, because, without radiative transfer 
calculations, we lack a precise association of our sink particles with protostellar classes, because the low-luminosity tail of our PLF is uncertain
also due to the simple modeling of the envelope luminosity (see the previous section), and because the low-luminosity tail of the PLF of 
\citet{Kryukova+2012} is uncertain due to their method of extrapolating the bolometric luminosity from the mid-IR luminosity.

\begin{figure*}[t]
\includegraphics[width=18cm]{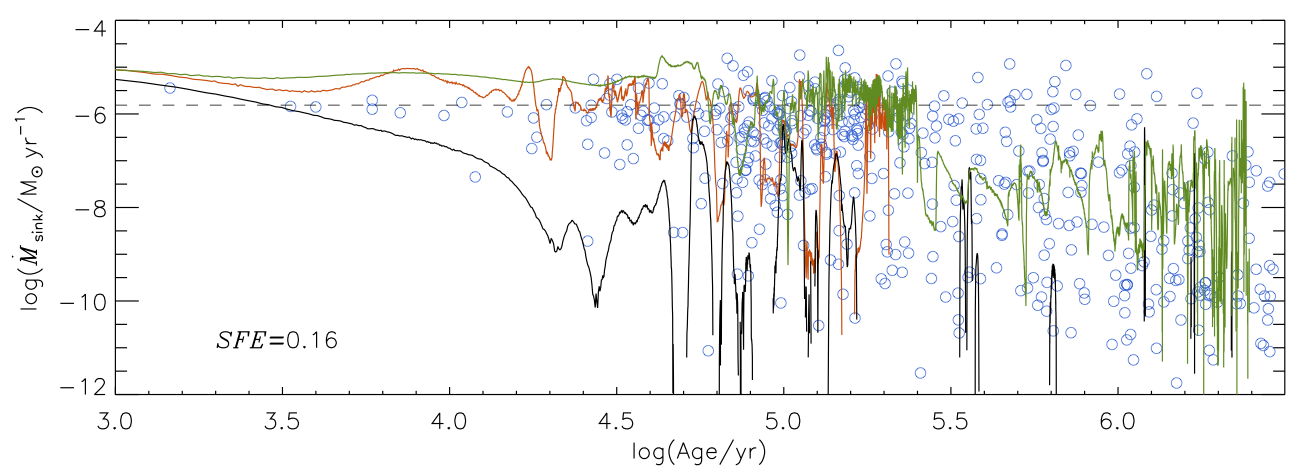}
\caption{Evolutionary tracks of infall rate versus age for three sinks showing large-amplitude fluctuations of the infall rate.}
\label{tracks_age}
\end{figure*}

\section{Discussion}

We have presented extensive evidence that our simulation yields protostellar luminosities matching the characteristic values of observed luminosities 
and their scatter in nearby star-forming regions. Recent works have addressed the problem of explaining the PLF either with simple analytical 
models, neglecting radiative transfer and the contribution of the envelope luminosity \citep[e.g.][]{Offner+McKee2011,Myers2012}, or based on
radiative transfer modeling of hydrodynamic simulations of protostellar collapse and disk evolution \citep[e.g.][]{Dunham+Vorobyov2012}. As discussed in 
\citet{Dunham+2013}, analytical models by  \citet{Offner+McKee2011} and  \citet{Myers2012} show that the luminosity problem 
may be solved by a scenario where the accretion time is independent of stellar mass (hence larger accretion rates for more massive stars). On the 
contrary, the solution proposed by \citet{Dunham+Vorobyov2012} is based on accretion rates that decrease with time and also experience a large 
number of high-amplitude bursts, as in the episodic-accretion scenario already envisioned by \citet{Kenyon+90} and \citet{Kenyon+Hartmann95}.  

Our work differs in a fundamental way from these previous studies. We have not tried to model nor to simulate the collapse of individual, isolated cores,
as that requires ad hoc assumptions to define initial and boundary conditions of individual objects. We rather stress the importance of the larger-scale
environment of protostellar cores, and the need to describe self-consistently the formation of a large number of them, in order to obtain a 
statistical distribution of realistic initial and boundary conditions for their formation. Protostellar cores are formed ab initio in our simulation, by a
process of turbulent fragmentation where we only control the large-scale mean parameters, such as the rms sonic and Alfv\'{e}nic Mach numbers
and the virial parameter, matched to their characteristic values in star-forming regions. 

Based on the extensive numerical literature on turbulent
fragmentation, we already know that simulations of driven, supersonic, MHD turbulence reproduce many observed properties of molecular clouds,
such as velocity and density scaling \cite[e.g.][]{Padoan+2003scaling,Heyer+Brunt04,Padoan+06perseus,Kritsuk+2013}, gas density distribution 
\citep[e.g.][]{Padoan+97ext,Padoan+99per,Schnee+2006,Brunt+2010,Price+2011,Kritsuk+2011,Kainulainen+Tan2013}, 
magnetic properties \citep[e.g.][]{Padoan+Nordlund99MHD,Lunttila+2008,Lunttila+2009,Heyer+Brunt12taurus}, 
cloud and core kinematics \citep[e.g.][]{Padoan+99per,Padoan+2001cores,Klessen+2005}.
Furthermore, the simulation of this work also yields a star formation rate and a complete stellar IMF in excellent agreement with observations.  

Cores that arise from the larger-scale turbulent dynamics of molecular clouds display fundamental features that may be quite different from 
those assumed in idealized models of single cores. Most notably, they are not isolated, not even at later evolutionary stages. The converging flows
responsible for its formation may persist for some time after a core has started to collapse into a protostar. This is in particular true for the stars formed 
in the densest environments in the model, and the massive stars in our simulation typically have converging flows persisting for more than a million years.
This results in a scenario where the infall rate may be dominated by converging flows from larger scales, rather than limited by the infall of a finite mass 
reservoir, whose value is determined at the very moment when the core becomes supercritical and starts to collapse. Our simulation shows that such 
infall rates control the formation of protostars, and allows us to quantify their statistical distribution, and thus the rate of growth of protostars and the 
resulting scatter in their luminosities.

Based on the evolution of the infall rates of sink particles in our simulation, we find that the luminosity problem is solved thanks to the general decline 
with time of the infall rates and to their large variations in time and from star to star, as originally envisioned by \citet{Kenyon+90} and 
\citet{Kenyon+Hartmann95}, and, more recently, by \citet{Dunham+Vorobyov2012}, based on hydrodynamical simulations of core collapse and 
disk evolution \citep{Vorobyov+Basu2005,Vorobyov+Basu2006,Vorobyov+Basu2010}. While these simulations start from an isolated supercritical 
core with ad hoc initial parameters such as mass, angular momentum, and density and temperature profiles, the statistical distributions of protostellar 
core properties and the time evolution and statistical distribution of the infall rates in our simulation are self-consistently determined by the larger-scale 
dynamics. Infall rates vary significantly between stars of similar current or final mass, as a result of the stochastic nature of turbulent flows, and time variations 
of large amplitude are often related to the binary nature of protostars, with the binary fraction and properties also arising ab initio in the simulation.

Although the accretion rate from the disk to the protostellar surface may exhibit somewhat different frequency and amplitude distributions than the infall
rate, due to modulation of the infall rate by disk instabilities, the growth of protostars must be controlled primarily by the infall rate, not by disk instabilities,
because disks are almost never a significant mass reservoir (except possibly in the very final stages of protostellar evolution, where most of the final
stellar mass is already in place anyway). Furthermore, the accretion-rate variability found in hydrodynamic simulations neglecting the core formation process and the crucial role
of magnetic fields \citep[e.g.][]{Vorobyov+Basu2010} is not necessarily a realistic representation of the accretion-rate variability of actual protostars. Only multi-scale
zoom in MHD simulations of star formation can self-consistently probe the variability of the accretion rate \citep{Nordlund+2014}.

\subsection{Evolution of Individual Sink Particles}

In \S4 we interpreted the scatter plots of infall rate versus sink age and sink mass on the basis of a simple scenario where sinks accrete for some time at a rate 
of the order of $10^{-5}$--$10^{-6}$~\msun \, yr$^{-1}$, and then gradually decay over time, with larger final stellar masses resulting from a longer time spent at the initial, 
higher rate. This scenario was illustrated in the top panels of Figure~\ref{rates_mass_age}, with idealized evolutionary tracks of protostars of different final masses.  

In Figure~\ref{tracks_age_mass}, we show the actual tracks of four characteristic sinks ending up with different final masses. Their infall rates are plotted versus
sink age and masses, on top of scatter plots such as those in the top panels of Figure~\ref{rates_mass_age}, but showing all the sink particles from the final snapshot 
of the simulation, at $t=3.2$~Myr and $SFE=0.16$. It is clear that the observed scatter in infall-rates (hence accretion luminosities) is the result of i) the decrease with time
of the infall rate of each sink, ii) large time fluctuations of the infall rate of individual sinks, and iii) differences in the infall-rate evolution between sinks with roughly equal final 
mass (orange and green tracks) or between sinks with very different final mass (all other tracks). 

One can also see that, after a brief period of rather large accretion rate, of the order of $10^{-5}$~\msun \,yr$^{-1}$, the infall rates may gradually decrease, though 
with very large fluctuations, when a low-mass star is formed, or remain at a relatively large value for an extended period of time ($\sim 1$~Myr), if a more massive 
star is to be formed, as in the idealized scenario presented in \S4. Such a prolonged formation time of massive stars should be accounted for in the computation of 
pre-main sequence stellar evolutionary tracks, particularly if the resulting isochrones are to be used to date young stars of a few solar masses \citep[e.g.][]{Hosokawa+2011}. 

To further quantify the formation timescale, we plot 
the age of individual sink particles when they have assembled 95\% of their final mass, versus their final mass, in Figure~\ref{age95}. We show all 631 sink particles
formed by $t=2.6$~Myr from the formation of the first one, and follow their evolution all the way to the end of the simulation, at $t=3.2$~Myr, at which point their mass
is assumed to be the final one. The large majority of these sink particles have indeed stopped growing by that time. The solid black line shows the result of a linear fit 
in the log-log plot, Age$_{95\%}=0.45\,{\rm Myr} \,(M_{\rm f}/{\rm M}_{\odot})^{0.56}$. In other words, it takes, on the average, 0.12, 0.45, and 1.63~Myr to form  
0.1, 1.0, 10.0~\msun\ stars, respectively. But the scatter plot also shows that stars of any mass may take up to 1--2~Myr to form, while stars are hardly ever formed with   
average infall rates much larger than $10^{-5}$~\msun \,yr$^{-1}$. The upper envelope on the formation time is most certainly artificially related to the limited time our 
simulation has been evolving, and longer simulation times are needed to reliably establish it. 

The lower envelope of the scatter plot (as most quantitative conclusion of this work) certainly depends 
on environment. In regions of massive star formation with an average density and rms velocity significantly larger than in our simulation, infall rates can be significantly 
larger than in our simulation (if infall rates are much in excess of $c_{\rm s}^3/G$, meaning that a mass larger than a critical one can be assembled by converging flows
in less than a free-fall time of the critical mass, then the maximum accretion rates are of the order of the free-fall time, irrespective of temperature, thus possibly
significantly larger than $c_{\rm s}^3/G$). In that case, the plot in Figure~\ref{age95} would be shifted to lower ages, the lower envelope would correspond to an 
infall rate larger than $10^{-5}$~\msun \,yr$^{-1}$, and stars much more massive than 10~\msun\ could be formed in 1-2~Myr.

\begin{figure}[t]
\includegraphics[width=8.7cm]{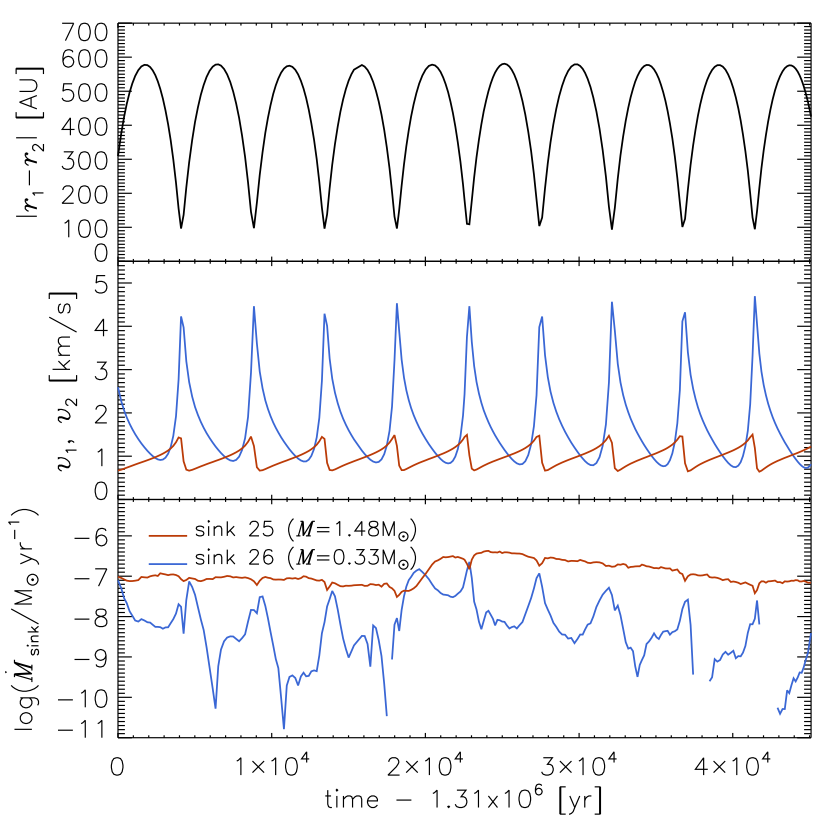}
\caption{Separation, velocities, and infall rate versus time, for a binary system composed of a 0.33 \msun\ sink orbiting around a primary of mass 1.48 \msun . The infall
rate of the secondary grows by 2-3 orders of magnitude at the approximate time of the periastro, to become comparable to the infall rate of the primary. }
\label{binary}
\end{figure}
\begin{figure*}[t]
\includegraphics[width=9cm]{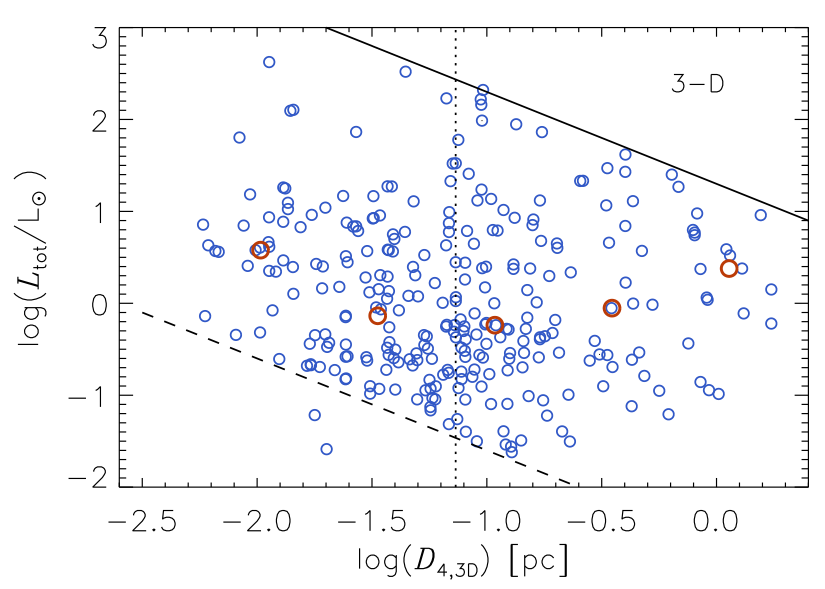}
\includegraphics[width=9cm]{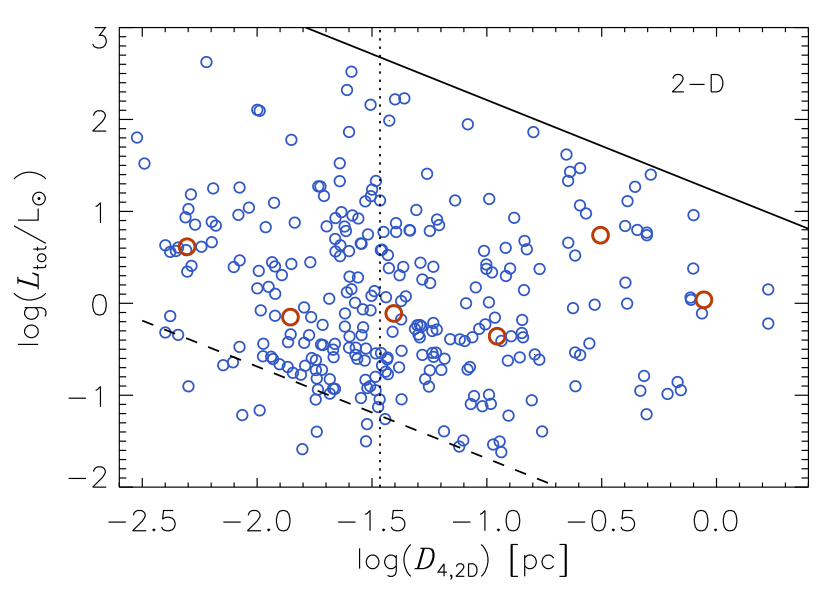}
\caption{Left Panel: Sink-particle total luminosity versus nearest-neighbor distance evaluated in 3D. The solid and dashed lines are approximate locations of the upper and 
lower envelopes of the plot, while the vertical dotted line shows the median value of $D_{4,3D}$. The median values of total luminosity computed in intervals of $D_{4,3D}$
is shown by the large, red, empty circles. Right Panel: Same as left panel, but with sink distances evaluated from a 2D projection.}
\label{d4}
\end{figure*}

In Figure~\ref{tracks_age}, we show the tracks of infall rate versus age for three other sinks, to illustrate the possibility of extreme fluctuations in infall rates. 
These particular sinks (like many others in our simulation) have infall rates that fluctuate many times between $\sim 10^{-6}$~\msun \,yr$^{-1}$ and 
$~10^{-11}$~\msun \,yr$^{-1}$ or even much lower values. This shows that even a single sink can in principle cover much of the observed luminosity scatter 
of protostars. However, a complete explanation of the luminosity scatter has to self-consistently include the formation of stars of all masses, yielding the correct
PLF as well as the observed stellar IMF. 

The apparent periodicity of the infall rate variations in many of our sinks is related to their binary nature. The largest fluctuations are found in the 
infall rate of the secondary member of a binary system, when the sink particle separation is of the order of a few hundreds AU. Figure~\ref{binary} 
shows an example of this, where a secondary sink of 0.33~\msun\ experiences periodic fluctuations in the infall rate of up to three orders of magnitude,
with the highest rate found at the periastro. At its smallest distance of approximately 100~AU from the primary, the secondary sink has an infall rate 
of the order of that of the primary, because it shares the very high density gas infalling on the primary.

\subsection{Turbulence, Core Collapse and Competitive Accretion}

\citet{Krumholz+2005} have contrasted two alternative scenarios of star formation, core collapse and competitive accretion, and have argued in favor of the former,
meaning that stellar masses are limited by the mass of prestellar cores. \citet{Bonnell+Bate2006} countered that competitive accretion is a viable explanation
for the origin of massive stars, if one accounts correctly for the cluster potential and the distribution of turbulent velocities and gas densities. 

It is often said that turbulent fragmentation predicts the mass function of prestellar cores, not the stellar IMF, and thus one has to assume a mass-independent 
core-to-star efficiency in order to derive the correct stellar IMF, hence turbulent fragmentation would imply a core-collapse picture. This is not the case for our scenario 
of turbulent fragmentation, which is neither core collapse, nor competitive accretion. We explained in \citet{Padoan+Nordlund2011imf} that our model \citep{Padoan+Nordlund02imf} 
yields a prediction for the stellar IMF, rather than the mass function of prestellar cores, because it estimates the total mass that the turbulence collects within a 
converging-flow region, not only the fraction of that mass collected into the core before the core starts to collapse. It follows directly from the model (from the turbulence velocity scaling) 
that more massive stars require converging flows of longer duration (larger scales) than lower mass stars, and thus their prestellar core mass (times a core-to-star efficiency), 
at the time when the collapse starts, can only be a fraction of their final mass. The larger the final stellar mass, the smaller this initial fraction. We also derived the mass function 
of prestellar cores based on our model, and showed that it must be significantly steeper than Salpeter's (Figure 4 in \citet{Padoan+Nordlund2011imf}), because massive 
stars must accrete much of their mass well after the beginning of the collapse of the prestellar core.

On the other hand, this extended accretion process for the more massive stars is different from competitive accretion, as it is directly driven by the same converging flow that
originated the prestellar core, rather than by the gravitational potential of the protostar, in competition with the potential of other stars or of the whole cluster (or
protocluster clump), as in the case of competitive accretion. This is why our IMF model \citep{Padoan+Nordlund02imf} predicts the Salpeter slope of the stellar IMF 
directly from the scaling of turbulence, without any reference to gravity (which is instead used in our prediction of the peak of the IMF and its shape around and below 
the peak), while the gravity of both individual protostars and the whole cluster (or protocluster clump) are crucial in deriving the IMF in the competitive accretion model 
\citep[e.g.][]{Bonnell+2001b}. 

However, the distinction between turbulent fragmentation and competitive accretion becomes more subtle in more realistic models and simulations,
where both gravity and turbulence must play a role. For example, in their analytical estimate of the accretion rate of massive stars, \citet{Bonnell+Bate2006} accounted for the fact 
that cores are formed by turbulent shocks, and that the gas velocity scales as in turbulent flows. Furthermore, numerical studies of competitive accretion do not completely
neglect the effect of turbulent fragmentation, as their initial condition includes a random velocity field with a realistic power spectrum 
\citep[e.g.][]{Bonnell+2003,Bate+Bonnell2005,Bate2009}, although to properly account for
turbulence, the turbulent flow would have to be driven until it is statistically relaxed, as we do in all our numerical studies of turbulent fragmentation (including this work). 
Likewise, the late infall rates of pre-main sequence stars (after they have left their parent converging-flow region) is of the Bondi-Hoyle type also in our simulation, and in regions
of high stellar densities such stars must be competing for the available gas mass (though most of their final mass is obtained in the earlier phase dominated by the local converging 
flow), and several low-mass stars end their growth because of a sudden ejection from multiple stellar systems. Aspects of both turbulent fragmentation and competitive accretion 
may eventually turn out to be important for a complete picture of star formation. 

Nevertheless, in the context of current analytical models of the PLF, there is no ambiguity in what is meant by competitive accretion. These models are based on specific assumptions
about the accretion rate of protostars. In the case representing competitive accretion, the protostellar mass is assumed to grow at a rate of ${\dot M}\propto M^j$ (where $M$ is the mass 
of the star, and ${\dot M}$ its growth rate), with for example $j=2/3$ in \citet{Offner+McKee2011}, $j=1.2$ in \citet{Myers2012}, or $j=1$ in \citet{Myers2014}, inspired by standard 
formulations of the competitive accretion, giving ${\dot M}\propto M^{2/3}$, in the case of gas-dominated potentials, or ${\dot M}\propto M^2$, in the case of stellar-dominated 
potentials \citep[e.g.][]{Bonnell+2001,Bonnell+2001b}.

The comparison of these analytical models with the observed PLF has led to the conclusion that competitive accretion is consistent with the PLF 
\citep{Offner+McKee2011,Myers2012,Myers2014}, or even the preferred star formation scenario \citep{Kryukova+2012}. We have shown in the previous 
subsection that Age$_{95\%} \propto M_{\rm f}^{0.56}$, so the time-averaged infall rate of our sink particles, $M_{\rm f}/$Age$_{95\%}\propto M_{\rm f}^{0.44}$, 
grows with final sink-particle mass, in rough agreement with the results of the analytical models assuming competitive accretion (though the lower envelope 
of the scatter plot in Figure \ref{age95} shows that the largest time-averaged infall rates are independent of final mass). However, those models assume that the
accretion rate of a given protostar grows with the protostar mass, while our evolutionary tracks for single sink particles never show such a trend. The characteristic, average evolution 
of the sink particles is to remain at relatively high infall-rate values of order $10^{-5}$--$10^{-6}$~\msun \,yr$^{-1}$ for a time roughly proportional to the final sink 
mass, and then to decline more or less rapidly once most of the final mass has been accreted, as illustrated by the idealized tracks depicted in the top panels of 
Figure~\ref{rates_mass_age}, and as shown in our plots of evolutionary tracks of actual sink particles. The complete absence of a dependence of the infall rate 
on sink mass during the evolution of individual sink particles demonstrates that competitive accretion, as implemented in those analytical models,
is not a viable explanation of the PLF. More realistic competitive accretion models, for example including density and velocity fields characteristic
of supersonic turbulence, cannot be ruled out at this stage based on the observed PLF.

The fact that the infall rates responsible for the formation of massive sinks remain at similar values when the sinks are very massive, as when they had a very low 
initial mass, indicates that the infall rates are controlled by the turbulent flow. The sink gravity is evidently too weak to significantly perturb the velocity field at a distance 
from the sink where most of its future mass reservoir resides. 

The observed PLF is also known to be biased towards higher luminosities in regions of higher stellar density \citep[e.g.][]{Kryukova+2012,Kryukova+2014,Elmegreen+2014}, 
which may be interpreted as due to mass segregation, possibly a manifestation of competitive accretion, or to larger accretion rates in regions of higher stellar density. 
The same trend is found in the PLF of our simulation. As a measure of protostellar density, we compute nearest-neighbor 
distances for the sink particles we have selected to compute the embedded PLF. Following \citet{Kryukova+2012}, we use the distance to the 4th nearest sink particle. 
We refer to it as $D_{4,3D}$ when it is computed using the three-dimensional (3D) coordinates of the sink particles, and $D_{4,2D}$, when it is computed in projection, 
using two-dimensional (2D) coordinates. The total luminosity of the sink particles is plotted versus their nearest-neighbor distance in Figure \ref{d4}.

The right panel of Figure \ref{d4} shows the 2D case, which can be directly compared with Figure 12 in \citet{Kryukova+2012}, except for the linear scale of  
$D_{4,2D}$ in their plots. As in the case of the high-mass star forming regions (and in some of the low-mass ones as well), our scatter plot has a rather well defined 
upper envelope, indicating a trend of increasing luminosity with increasing stellar density, for the highest luminosity sources. We have verified that such a trend is
at least in part due to a trend of increasing sink mass with increasing stellar density, though the upper envelope has also a contribution from increasing infall
rates with increasing stellar density, particularly for $D_{4,2D}>0.1$ pc. We thus confirm a certain amount of initial mass segregation among protostars. 
However, the mass segregation is only (and partly) manifested by the upper envelope. The median values of the logarithm 
of the total luminosities, computed in logarithmic $D_{4,2D}$ intervals (large red circles in the right panel of Figure \ref{d4}) do not show any clear trend with 
stellar density. 

The result is confirmed with 3D nearest-neighbor distances, shown in the left panel of Figure \ref{d4}. The upper envelope is even better defined than in 2D, 
but, nevertheless, the median values of total luminosity do not follow a monotonic trend with $D_{4,3D}$. The median luminosity has a minimum at 
$D_{4,3D}\approx0.1$ pc, and increases both towards lower and higher values of $D_{4,3D}$. Furthermore, the rather well-defined lower envelope has
nothing to do with mass segregation, as it is completely controlled by the envelope luminosity. If we removed the envelope luminosity, the lower envelope 
of the scatter plot would not be well defined any more, and the minimum values of luminosity would drop significantly. Thus the increase of the median luminosity 
with decreasing $D_{4,3D}$ is not indicative of mass segregation, as it is mainly due to the lower envelope of the scatter plot. 

In summary, we find a trend in the upper envelope of the scatter plot of total luminosity versus nearest-neighbor distance consistent with that from
high-mass star forming regions, but that trend is barely an indication of mass segregation. There is a tendency for massive stars to be found in regions 
of high stellar density, but in those regions one always find also stars of lower masses. Future work should test if this low level of mass segregation 
is consistent with that predicted when competitive accretion is the dominant formation mechanism of massive stars.


\section{Summary and Conclusions}

We have proposed a new paradigm where infall rates in a turbulent cloud control the protostellar luminosity, and drive the disk accretion process even after the end of
the embedded protostellar phase. After showing that our simulation yields a realistic star formation rate and a complete stellar IMF consistent with the observations, 
we have demonstrated that also the infall rates from the simulation are consistent with the observations, under the reasonable assumption that the infalling 
gas finds its way to the stellar surface, irrespective of the specific disk process that allows this to occur. 

Besides offering an appealing interpretation of the observed protostellar accretion rates, this new paradigm also naturally solves the luminosity problem, as the
infall rates in a turbulent cloud result in protostellar luminosities that match both the observed characteristic values and their scatter. Our simulation also reproduces
the observed PLF. Our main findings are summarized as follows.

\begin{itemize}

\item Infall rates controlled by converging flows in turbulent clouds are of the same order of magnitude as accretion rates inferred from protostellar luminosities
and estimated accretion rates of young PMS stars. The growth and accretion luminosity of protostars, as well as the accretion rate of young PMS stars are thus 
primarily controlled by such infall rates. 

\item Scatter plots of infall rate versus sink particle mass and age, as well as the direct examination of the time evolution of individual sink particles, show
that, on the average, the infall rates remain relatively high, $\sim 10^{-5}$--$10^{-6}$~\msun \,yr$^{-1}$, for a period of time that is longer for larger final
masses, and then decrease with time. Individual sink particles experience very large fluctuations from this average behavior, due to the stochastic nature
of the converging flows in the turbulent velocity field of star-forming clouds.  

\item The scatter plot of infall rate versus mass has an upper envelope that scales approximately as $\propto M^2$, once all sink particles younger than 
approximately 0.5 Myr (or more) are removed, in agreement with the upper envelope of observed PMS-star accretion rates versus mass. However, this 
trend should not be viewed as a direct constraint on the time evolution of individual PMS stars, as the accretion rate of individual sinks does not appear to
systematically grow over time as the sink mass increases.

\item Infall rates remain significant, meaning of the order of observed accretion rates of PMS stars, for the whole duration of our simulation, 3.2 Myr after the formation 
of the first sink particle, for a large majority of the sinks. Mass infall is thus important even beyond the embedded protostellar phase, and we caution
against assuming that most disks of Class II (and possibly even Class III) sources evolve as isolated systems controlled only by internal processes. Even when 
sink particles have left their birth site, a converging region in the turbulent flow, and have already acquired most of their final mass, Bondi-Hoyle accretion can 
account for the observed accretion rates.

\item The stochastic nature of infall rates controlled by the cloud turbulence results in accretion luminosities covering a total scatter of many orders of magnitudes,
for sink particles of the same age. The observed scatter of 2--3 orders of magnitude in the total luminosity of protostars is thus naturally reproduced in the simulation.
The characteristic values of total luminosity we derive are also consistent with the observations. Thus, once realistic protostellar infall rates are accounted for, 
the luminosity problem is solved. Our results support the existence of episodic accretion events as a consequence of wide variability in the infall rates, though 
the modulation by disk physics should be included for a detailed comparison with observational evidence of episodic accretion.

\item The accretion luminosity is correlated with the envelope mass measured at a fixed radius of 2,500 AU, $M_{2500}$. This leaves a visible imprint in the 
scatter plot of total luminosity versus $M_{2500}$, which shows a trend towards larger luminosities with increasing  $M_{2500}$. This plot allows a rather
direct comparison with the observations, as it does not rely on protostellar class and does not require a very detailed modeling of the spectral energy distribution.

\item The observed protostellar luminosity function (PLF) is also nicely reproduced. However, one should not expect a perfect match due to the dependence of
the PLF on the environment, for example stellar density, and due to rather large uncertainties in both modeling and observing the low-luminosity tail of the PLF. 

\item The luminosity problem, meaning both the low characteristic protostellar luminosity and the large scatter from protostar to protostar, is solved through the
contribution of at least three aspects of infall rates in turbulent clouds:  i) the decrease with time of the infall rate of each sink particle, ii) the occurrence of time 
fluctuations of large amplitude in the infall rate of individual sink particles, and iii) the very large differences in the infall-rate evolution between sinks, both in the 
case of similar or very different final masses. 

\item Although the time-averaged infall rate increases with the final sink mass, as in the analytical models of the PLF assuming competitive accretion, infall rates of 
individual sink particles do not increase systematically as the sink mass increases, in direct contradiction with the specific assumption of those 
models. This shows that the gravitational force of a star is too weak to shape the velocity field of infalling gas at a distance from the star where most of its future mass 
reservoir resides. Infall rates are thus a feature of the cloud turbulence that spontaneously sets regions of converging flows. 

\item From the viewpoint of the evolution of individual protostars, the solution of the luminosity problem under this turbulent fragmentation scenario is similar to
that of the episodic accretion model of \citet{Dunham+Vorobyov2012}, as it relies on both time-decay and large-amplitude fluctuations 
of the accretion rate, as already envisioned by \citet{Kenyon+90} and \citet{Kenyon+Hartmann95}. However, our solution is still fundamentally different 
from that of \citet{Dunham+Vorobyov2012}, in that it is the time variation of the infall rates controlled by the turbulent flow that drives the time-decay of the
accretion rate and the episodic events, rather than the infall from the collapse of an isolated core and disk instabilities. 

\item Besides the stochastic nature of converging flows in the cloud turbulence, the observed luminosity scatter may hide a contribution from the binary nature 
of many protostars, as we find that binary sink particles often exhibit large fluctuations of their infall rates, strongly correlated with their orbital period. 

\end{itemize}

A definitive physical explanation of the luminosity problem must ultimately address the basic question of why we find embedded sinks (protostars) with low 
infall (accretion) rates. The termination of protostellar growth may take several paths, such as the exhaustion of the converging flow, the gradual displacement 
of such flow from the protostar due to random magnetic and ram pressure forces in the turbulence (with some possible contributions from stellar feedbacks) 
acting on the gas and not on the protostar, the sudden displacement of the protostar from the birth site due to dynamical interactions with other stars. 
All of these processes play a role, and future work should quantify their relative importance.

\acknowledgements
We thank Carlo Manara and Leonardo Testi for providing the data of Figure 5 (Manara et al. 2014) and for many useful explanations 
about observational measurements of accretion rates. We also acknowledge helpful discussions with Tom Megeath, Mike Dunham, Lee Hartmann, 
Neal Evans, and other participants in the Oort Workshop ``Episodic Accretion", held at the Lorentz Center, University of Leiden, on May 13th-15th, 2014. 
Mike Dumham, Neal Evans, Lee Hartmann, Carlo Manara, Chris McKee, Tom Megeath, Stella Offner, and the anonymous referee provided useful comments on the
first version of the manuscript. 
Computing resources for this work were provided by the NASA High-End Computing (HEC) Program through 
the NASA Advanced Supercomputing (NAS) Division at Ames Research Center, and by the Port d'Informaci\'{o} Cient\'{i}fica 
(PIC), Spain, maintained by a collaboration of the Institut de F\'{i}sica d'Altes Energies (IFAE) and the Centro de Investigaciones 
Energ\'{e}ticas, Medioambientales y Tecnol\'{o}gicas (CIEMAT). PP acknowledges support by the FP7-PEOPLE- 2010- RG grant 
PIRG07-GA-2010-261359. TH is supported by a Sapere Aude Starting Grant from The Danish Council for Independent Research. 
Research at Center for Star and Planet Formation was funded by the Danish National Research Foundation and the University
of Copenhagen's programme of excellence.

\bibliographystyle{apj}

\end{document}